\documentclass[conference]{IEEEtran}
\usepackage[utf8]{inputenc}
\usepackage[T1]{fontenc}
\usepackage[english]{babel}

\usepackage{subcaption}
\usepackage{comment}
\usepackage[linesnumbered,lined,boxed,ruled,longend,procnumbered]{algorithm2e} % commentsnumbered,

\SetCommentSty{mycommfont}
\usepackage[amssymb,Gray,thinspace]{SIunits}
\addunit\decibelm{dBm}

\usepackage{amsmath}
\usepackage{amsfonts}

\usepackage{graphicx, float,  blindtext}

\newcommand\IEEEhyperrefsetup{
bookmarks=true,bookmarksnumbered=true,%
colorlinks=true,linkcolor={black},citecolor={black},urlcolor={black}%
}

\usepackage[\IEEEhyperrefsetup]{hyperref}

\usepackage{mathtools}

\usepackage[nolist,nohyperlinks]{acronym}
\usepackage{cleveref}
\graphicspath{{images/}}

\usepackage{titlesec}
\titlespacing*{\section}{0pt}{*1}{*1}
\titlespacing*{\subsection}{0pt}{*1}{*1}
\renewcommand{\thesubsubsection}{\arabic{subsubsection}}
\titleformat{\subsubsection}[runin]{\itshape}{\thesubsubsection)}{1em}{}[:]
\titlespacing*{\subsubsection}{\parindent}{0pt}{*1}

\author{\IEEEauthorblockN{Stefan Gvozdenovic\IEEEauthorrefmark{1},
Johannes K Becker\IEEEauthorrefmark{2}, John Mikulskis\IEEEauthorrefmark{3} and
David Starobinski\IEEEauthorrefmark{4}}
\IEEEauthorblockA{Department of Electrical and Computer Engineering, Boston University\\
Boston, MA 02215\\
Email: \IEEEauthorrefmark{1}tesla@bu.edu,
\IEEEauthorrefmark{2}jkbecker@bu.edu,
\IEEEauthorrefmark{3}jkulskis@bu.edu,
\IEEEauthorrefmark{4}staro@bu.edu}}

\title{IoT-Scan: \\ Network Reconnaissance for the Internet of Things}
\begin{document}
% Create the title.
\maketitle
\begin{abstract}
Network reconnaissance is a core networking and security procedure aimed at discovering devices and their properties. For IP-based networks, several network reconnaissance tools are available, such as Nmap. For the Internet of Things (IoT), there is currently no similar tool capable of discovering devices across multiple protocols. In this paper, we present  {\tt  IoT-Scan}, a universal IoT network reconnaissance tool.
 {\tt  IoT-Scan} is based on software defined radio (SDR) technology, which allows for a flexible software-based implementation of radio protocols. We present a series of 
passive, active, multi-channel, and multi-protocol scanning algorithms to speed up the discovery of devices with  {\tt  IoT-Scan}. We benchmark the passive scanning algorithms against a theoretical traffic model based on the non-uniform coupon collector problem.  We implement the scanning algorithms and compare their performance for four popular IoT protocols: Zigbee, Bluetooth LE, Z-Wave, and LoRa. Through extensive experiments with dozens of IoT devices, we demonstrate that our implementation experiences minimal packet losses and achieves performance near the theoretical benchmark. Using multi-protocol scanning, we further demonstrate a reduction of 70\% in the discovery times of Bluetooth and Zigbee devices in the 2.4\,GHz band and of LoRa and Z-Wave devices in the 900\,MHz band, compared to sequential passive scanning.  
We make our implementation and data available to the research community to allow independent replication of our results and facilitate further development of the tool.
\end{abstract}
% Example sections, name them
% according to specific needs.
\section{Introduction}

% Highlight motivation, current state of the art, and needs.
% scanning non-IP based protocols
% single hardware platform
% future-proof (new IoT protocol, new version)
% Parallel scanning vs sequential (multichannel multiprotocol)

\begin{comment}
Digital wireless communication is experiencing rapid growth driven by ever-expanding consumer and commercial applications around the Internet of Things (IoT). IoT networks come in different flavors. Most of the devices we tested fall under the consumer smart home and wearable device category. Wearable devices often use peer-to-peer protocols (e.g. BLE). Smart home networks types tend to be star shaped (e.g. Zigbee), or mesh (Z-Wave). Mesh network can span one's private network (Z-Wave smart home security) or can hop across neighboring IoT gateways (e.g. LoRa, Sidewalk~\cite{amazon-sidewalk}). Consumer IoT networks often include several end devices (sensors, actuators, etc) that are connected to a gateway(s), which is connected to the internet (via Wi-Fi or Ethernet). Usually, an IoT gateway is specific to an IoT protocol, although some support multiple IoT protocols (e.g. Echo, Ring base station).
\end{comment}

The Internet of Things (IoT) device market is currently exhibiting exponential growth (among the 29 billion connected devices forecast this year, 18 billion will be related to IoT~\cite{ericsson}). These devices run a variety of low-power communication protocols, such as Bluetooth Low Energy (BLE)~\cite{BluetoothCoreSpec5}, Zigbee~\cite{ieee802.15.4}, Z-wave~\cite{9959}, and LoRa~\cite{Tapparel2020}, which support applications in smart homes, smart grid, health care, and environmental monitoring. 

The heterogeneity of the IoT ecosystem -- and in particular the large number of  IoT protocols -- represents a major challenge from a network and security monitoring perspective~\cite{Ortiz2019,Huang2020IoTInspector}. This heterogeneity makes it hard for network administrators to run network reconnaissance tasks, which aim at discovering wireless IoT devices and their properties. Network reconnaissance is crucial for maintaining an asset inventory, monitoring changes in device behavior, and detecting rogue devices. Since many IoT devices are mobile (e.g., wearables and trackers), network reconnaissance tasks must be run regularly.  New laws adopted by regulators, such as the IoT Cybersecurity Improvement Act of 2020~\cite{IoT-Act} in the US, provide further impetus to the design of effective solutions for IoT network reconnaissance.
 
The simplest solution for IoT network reconnaissance is to use a monitoring device equipped with a different network card for each protocol. However, even devices operating on the same protocol may be incompatible if they run different versions of the protocol (e.g., normal versus long-range Z-Wave~\cite{zwave-longrange}). 
%Additionally, some protocols are proprietary, such as LoRa~\cite{Tapparel2020}. 
Using dozens of different USB dongles or network cards for each protocol is prohibitive for practical network security auditing.  

Existing software tools for network reconnaissance, such as Nmap~\cite{lyon2008nmap}, focus on devices with IP addresses. Nmap can scan IP/port ranges for an arbitrary number of local or remote hosts and their services. However, this approach is limited to IP-enabled devices only, and yet many popular IoT protocols, including BLE, Zigbee, Z-Wave, and LoRa do not support IP addressing. While there is currently an effort by several vendors to create a unified, IP-based IoT protocol, called Matter~\cite{project-matter_2019}, its level of adoption and  backward-compatibility with legacy devices remain uncertain.

To address this current gap, we propose {\tt IoT-Scan}, an extensible, multi-protocol IoT network reconnaissance tool for enumerating IoT devices. {\tt  IoT-Scan} runs both on the 900\,MHz and 2.4\,GHz bands and currently supports four popular IoT protocols: Zigbee, BLE, LoRa, and  Z-Wave. Remarkably, {\tt IoT-Scan} runs on a single piece of hardware, namely a software-defined radio (SDR)~\cite{ulversoy2010software}.

 {\tt  IoT-Scan} leverages software-defined implementation of IoT communication protocol stacks, mostly under the GNU Radio ecosystem~\cite{gr,scapy-radio}. This approach reduces the amount of hardware needed to address the growing number of IoT protocols. This further allows for future expansion into
 %into other IoT protocols and 
 new protocol versions, thus eliminating the need of purchasing or upgrading protocol-specific  hardware~\cite{He2010MPAP}.
 %Granelli2015,Li2018}. %Addressing multiple protocols with common hardware via a virtualized physical-layer architecture has been proposed as a solution to the growing number of protocols throughout the last decade~\cite{He2010MPAP,Tan2012,Granelli2015,Li2018}.
%In this work, we demonstrate SDR-based scanner for device discovery across multiple protocols on the 2.4GHz and sub-gigahertz %(900MHz in USA)
%ISM bands using a single SDR device.

%applies to several 

%can effectively discover devices both on the 900\,MHz band,  {\tt {\tt IoT-Scan}} runs on a single software-defined radio (SDR) platform.

A key challenge faced in the design of {\tt IoT-Scan} lies in minimizing the discovery time of devices. A simple approach (which we refer to as a \emph{sequential} algorithm) is to scan devices in a round robin fashion across each individual protocol, and in turn across each individual channel within each protocol. However, this approach does not scale. Consider, for instance, that Zigbee devices can communicate over 16 different channels.

To address the above, we propose, implement, and benchmark several scanning algorithms to speed up the discovery of IoT devices. These algorithms, of increasing sophistication, can listen in parallel \emph{across different channels and different protocols}. To achieve this, our work takes on the challenge of integrating single-protocol physical layer implementations of IoT protocols for SDRs to perform parallel scanning across channels and protocols, under the constraints of limited instantaneous bandwidth (i.e., the range of frequencies to which the SDR is tuned at a given point in time). Indeed, the channel spread defined by most protocols operating in the 2.4~GHz band is wider than the typical instantaneous bandwidth of an SDR, i.e., it is typically not possible to monitor the entire spectrum of a protocol simultaneously with one monitoring device. 

Another challenge is that some IoT devices transmit sparingly (e.g., due to energy savings considerations), and enumerating devices passively on the wireless channel can result in long discovery times. To speed up discovery of such devices, we propose \emph{active scanning algorithms} that send probe messages to discover which channels are actively used by devices of a given protocol, and skip channels on which no communication is taking place. We demonstrate these algorithms for Zigbee devices.

Another important consideration is evaluating the efficiency of the scanning algorithm implementation on the SDR, namely whether devices are indeed discovered as fast as possible and no packet loss is incurred due to imperfect SDR implementation. 
%If a scanning algorithm approaches the discovery times of a theoretical benchmark, then its performance is provably close to optimal. 
We achieve this by establishing a connection between our network scanning problem and the non-uniform coupon collector problem~\cite{Flajolet1992Coupon,Anceaume2015}, whereby each transmission by a specific device corresponds to a coupon of a certain type and the objective is to collect a coupon of each type as fast as possible. The non-uniformity of the problem stems from the different rates at which different devices transmit packets. Under appropriate statistical assumptions, we can analyze this problem and numerically compute the expectation of the order statistics of the discovery times (i.e., the average time to discover $n$ out of $N$ devices, for any $n =1, 2, \dots, N$). For the cases of Zigbee and BLE, we show that the discovery times, as measured through several repeated experiments, closely align with these theoretical benchmarks.

Our main contributions are thus as follows:
%\vspace{-17pt}
\begin{itemize}
%(that take advantage of SDR, protocol freq. allocation)
    \item We introduce  {\tt IoT-Scan}, a universal tool for IoT network reconnaissance.  {\tt IoT-Scan} consists both of  a collection of efficient and practical IoT scanning algorithms and of their implementations using a single commercial off-the-shelf software-defined radio device, namely a USRP B200 SDR~\cite{ettus-usrp-b200}.
    \item We validate the performance of the algorithms through extensive experiments on a large collection of devices. We demonstrate multi-protocol, multi-channel scanning both on the 2.4~GHz band for Zigbee and BLE, and on the 900\,MHz band for LoRa and Z-Wave. %We show that these algorithms reduce the total scanning time by about a factor of 4 compared to a sequential algorithm.
    \item We propose new active scanning algorithms and show an implementation for Zigbee, which cuts down the discovery time by 87\%, from 365~s to 46~s,  compared to a sequential scanning algorithm.
    \item We develop a theoretical benchmark based on the non-uniform coupon collector problem, and show that passive scan algorithms for Zigbee and BLE achieves performance near that benchmark
    %\item We demonstrate the implementation of active multiprotocol scanning, which results in over 3 times faster device discovery times than the corresponding passive sequential scan.
    \item We discuss implementation challenges and parameter optimizations as they relate to scanning performance.
    %\item Efficient and practical parallel multi-protocol IoT scanning algorithms
%\item Improved single-channel single-protocol receivers for XXX protocols
% \item (Implementation of) Multi-channel SDR receiver for 802.15.4/Z-Wave/BLE/LoRa protocols
% \item (Implementation of) Multi-protocol SDR receiver for (several combinations of) said protocols 
\end{itemize}
%\vspace{-10pt}
The rest of this paper is structured as follows. 
Section~\ref{sec:related} discusses related work.
Section~\ref{sec:algorithms} presents the scanning methods and algorithms forming the core of {\tt IoT-Scan}. 
Section~\ref{sec:problem} discusses performance metrics for the algorithms, as well as a theoretical model for benchmarking device discovery.
Section~\ref{sec:background} provides background on each of the IoT protocols covered in this paper, and elaborates on how {\tt IoT-Scan} discovers addresses of devices in each case.
  Section~\ref{sec:exp} presents our experiments, including
%Section~\ref{sec:scanning} presents the scanning methods and algorithms
implementation aspects, experimental setup, and results.
Section~\ref{sec:conclusion} concludes our findings, discusses ethical issues, and presents an outlook on future work. %Appendix~\ref{sec:active_multi_proto} lists pseudo-code for one of the advanced scanning algorithms.

%\color{red}
%\begin{verbatim}
%III.    Protocol Device Enumeration
%    -   with some examples
%    -   List of devices in this sections so
%        subsequent sections can refer to it
%IV.     Problem Statement and Analysis
%V.      Scanning Methods
%VI.     Experimental Evaluation
%    a. Set-Up
%    b. Results
%VII.    Limitations
%\end{verbatim}
%\color{black}
\section{Related Work}
\label{sec:related}

This section presents related work. Most existing work focuses on \emph{protocol-specific} techniques. In contrast our work introduces several \emph{cross-protocol} algorithms for IoT scanning, and further benchmarks their performance both theoretically and experimentally.

Heinrich et al. presents BTLEmap~\cite{heinrich2020btlemap}, a BLE-focused device enumeration and service discovery tool inspired by traditional network scanning tools like Nmap~\cite{lyon2008nmap}. Aside from an extended message dissector based on both the BLE specification as well as previous reverse-engineering work on Apple-specific message types, BTLEmap shares similarities to tools available in the Linux Bluetooth stack. While BTLEMap supports both Apple's Core Bluetooth protocol stack and external scanner sources, it is limited to Bluetooth LE by design and does not aim to support multiple protocols. In contrast, {\tt IoT-Scan} is not tied to a particular vendor as a host device, and supports multiple protocols simultaneously, with one radio source.

Tournier et al. propose IoTMap~\cite{Tournier2020}, which models inter-connected IoT networks using various protocols, and deduces network characteristics on multiple layers of the respective protocol stacks. IoTMap has a strong focus on modeling network layers in a cross-protocol environment, as well as  identification of application behavior and network graphs across protocols. However, IoTMap requires dedicated radios for each protocol in order to operate, whereas {\tt IoT-Scan} achieves device detection across multiple protocols with a single software-defined radio transceiver.
%Mikulskis et al.

In a preliminary poster~\cite{snout},
we introduced a precursor to {\tt IoT-Scan} that showcased scanning of BLE and Zigbee devices with an SDR platform.
%Snout leverages GNU Radio as a platform to interact with multiple IoT protocols via SDR hardware.  
{\tt IoT-Scan} encompasses additional  protocols, namely LoRa and Z-Wave. Furthermore, our work introduces novel scanning algorithms and conducts extensive evaluation of these algorithms, both theoretically and empirically with dozens of IoT devices. In contrast, our preliminary work did not present scanning algorithms and had no evaluation contents (either theoretical or empirical).

%{\tt IoT-Scan} significantly expands beyond this previous work in terms of the number of protocols addressed (i.e., it also includes Z-Wave and LoRa).  and finally, demonstrates device enumeration scans in close alignment with the theoretical model, as well as simultaneous multi-protocol scanning. 
%inspired by this approach, but significantly expands  similarly targets wireless non-IP based IoT devices. It is extensible in theory as it uses SDR. However, it is currently limited to basic single-channel passive scans of two protocols: Zigbee and BLE. On the other hand, {\tt IoT-Scan} supports four protocols Zigbee, BLE, ZWave and LoRa. Including passive, active, and parallel multiprotocol scanning algorithms which are benchmarked to achieve best possible performance.

Bak et al. \cite{bak2019designing} optimize BLE advertising scan (i.e., device discovery) by using three identical BLE dongles. This approach is not scalable since it requires a new hardware receiver for each new channel, and equally does not scale beyond the BLE protocol.
%Furthermore, Ubertooth One hardware receivers are BLE specific.
In contrast, our SDR-based approach uses the same SDR hardware to receive multiple protocols.

Kilgour~\cite{kilgour2013bluetooth} presents a multi-channel BLE capture and analysis tool implemented on a field programmable gate array (FPGA). This multi-channel BLE tool allows receiving data from multiple channels in parallel. However, the focus is on BLE PHY receiver implementation and related signal processing rather than actual scanning and enumeration of devices. Kilgour's work discusses FPGA extensions for the USRP N210 platform which in theory allow for a large number of Bluetooth LE channels to be received in parallel. However, no practical validation is performed to demonstrate this configuration. In contrast, our work extends beyond Bluetooth LE, and crucially performs practical device enumeration scans to quantify scanning performance.

Active scan is a known device discovery technique used in Wi-Fi~\cite{wifiactive}. Thus, Park et al. describe a Wi-Fi active scan technique performed using BLE radio using cross-protocol interference \cite{park2021bless}. The active scan algorithms in {\tt IoT-Scan} are motivated by similar ideas, but require judicious use of protocol-specific mechanisms (i.e., sending beacon request packets in Zigbee).  
%In fact, Wi-Fi distinguishes between: passive, active, and directed scans. Unlike active scan which broadcasts probe requests and listens to probe responses, a directed scan uses unicast probes instead. 

Hall et al.~\cite{hall2016z} describe a tool, called EZ-Wave that can discover Z-Wave devices passively and actively.
%, that implements active scanning of Z-Wave devices built on top of scapy-radio~\cite{scapy-radio}.
The EZ-Wave tool actively scans a Z-Wave device by sending a ``probe'' packet with acknowledgement request flag set. In the older version S0 of the Z-Wave protocol, it was compulsory for a Z-Wave device to reply with acknowledgements to such packets. By getting this acknowledgement back, the EZ-Wave tool learns about a device's presence.
%Note that for a ``probe'' packet to be received by devices, it needs to have the correct home id (a network specific ID for Z-Wave), source/destination ID, sequence number which can all be sniffed with passive scan.
% etc. By adding the acknowledgement flag to the probe packet any responding non-battery powered Z-Wave device %(only Ring Security Station in our devices) is required to reply with an acknowledgement frame, revealing its source ID.
However, the EZ-Wave tool only supports older versions of Z-Wave protocol. In the new version (S2) of the Z-Wave protocol, acknowledgements are not compulsory and this is not a reliable active scan mechanism. The old Z-Wave protocol uses only the R1 (9.6\,kbps) and R2 (40\,kbps) physical layers. Our work adds R3 (100\,kbps PHY) as well as multi-protocol capabilities. The R1, R2, and R3 rates are defined in~\cite[Table 7-2]{9959}.
%It should be noted that the type of %Z-Wave active scan used by EZ-Wave does not work on Ring Z-Wave devices.
%S2 (Security 2) Z-Wave is more secure and energy efficient upgrade from legacy S0.
%\color{red}Even if such Z-Wave active scan was feasible we describe in Section III that it would not be significantly advantageous compared to passive scan (see Section III on Z-Wave).\color{black}

%zigbee ble lora scanners
Choong~\cite{choong2009multi} implement a multi-channel IEEE~802.15.4 receiver using a USRP2 software-defined radio.
The USRP2 has a maximum sample window of 25\,MHz and maximum Ethernet backbone (radio to PC communication) transfer rate of 30\,MS/s. This limits the multi-channel receiver to a maximum of five (consecutive) Zigbee channels. Choong describes a channelization  method similar to the receive chain used in this work (see Section~\ref{sec:signal-processing}) that extracts multiple channels from a wider raw signal stream. However, Choong's work focuses on the performance impact of the SDR host computer, and is a Zigbee-specific implementation, whereas our work focuses on device enumeration in a multi-channel as well as multi-protocol context.

Our Zigbee, BLE, and Z-Wave GNU Radio receiver implementations are based on scapy-radio \cite{scapy-radio} flowgraphs. Our LoRa GNU Radio receiver flowgraph is based on a work by Tapparel et al.~\cite{Tapparel2020}.
A similar multi-channel LoRa receiver was implemented by Robyns in \cite{robyns2018multi}.
In order to support multi-radio, multi-channel capabilities, {\tt IoT-Scan} implements several changes to these GNU Radio receiver implementations. In general, these changes pertain to the signal path between the SDR source and the receive chains for individual channels and protocols (i.e., frequency translation, filtering, and resampling, see Section~\ref{sec:algo-imp}). Additionally, our LoRa receiver can listen to LoRa packets promiscuously.
\section{Scanning Algorithms}
\label{sec:algorithms}

In this section, we introduce SDR-based scanning algorithms that form the core of {\tt IoT-Scan}. Alongside, we introduce several auxiliary helper functions. 
%The algorithms are described in a generic fashion, while 
%Sections~\ref{sec:exp} and Appendix~\ref{sec:background}
%discuss implementation details of these algorithms in the context of various specific IoT protocols.    
%We start by introducing a single-channel, single-protocol passive scan (Algorithm~\ref{alg:passive}), followed by an active version (Algorithm~\ref{alg:active}). We then propose a multi-channel, multi-protocol passive scan (Algorithm~\ref{alg:multiproto}), followed by an active version (Algorithm~\ref{alg:active-multiproto}). Alongside, we introduce several auxiliary helper functions (Functions~\ref{alg:listen},\ref{alg:probech},\ref{alg:chinrange}). %Implementation details on SDR are part of the next section.
The notion of \emph{channel} in this section refers to a 3-tuple containing the center frequency of the channel, the channel bandwidth (i.e., a range of frequencies delineated by the lower and upper frequencies of the channel), and the protocol type. The concept of \emph{instantaneous bandwidth} refers to the range of frequencies captured by the SDR at any given point of time. The \emph{center frequency} corresponds to the frequency at the middle of the range.

%\color{red}TODO: define SDR-specific things like center frequency, instantaneous bandwidth / channel bandwith, etc\color{black}

\subsection{Single-channel methods}

% Algorithm 1: Listen
\begin{algorithm}[tb]
\caption{Listen($ch,dwell\_time$)}
\label{alg:listen}
%\begin{algorithmic}[1]
% \Require $n \geq 0$
% \Ensure $y = x^n$
\SetKwComment{Comment}{$\triangleright$\ }{}
%\textbf{procedure} LISTEN($ch,dwell\_time$)\\
\Comment{Receive packets on channel $ch$ for duration $dwell\_time$ and return a list of discovered devices}
%\Comment{A channel $ch$ is characterized by a tuple ($freq, bw, proto$), i.e., its frequency $ch.freq$, channel bandwidth $ch.bw$ and protocol $ch.proto$.}
%\Procedure{Listen}{$ch,dwell\_time$}\\
$t_{start} \gets $time() \Comment*[f]{Store current time}\\
$device\_list \gets\{\}$ \Comment*[f]{Initialize device list}\\
\While{\textnormal{time()}$-t_{start} \le dwell\_time$}{
Listen on channel $ch$ \\
Get packet and extract address $dev\_addr$  \label{alg:listen:empty}\\
%\Comment{Note that if packet does not contain a device address, $dev\_addr$ is an empty element.} 
$device\_list = device\_list \cup dev\_addr$ \label{alg:listen:add-dev}\\
%\Comment{Add device to list.} 
%{Note that if device is already an element in the list, union operation does not change list.}  
}
%\EndWhile
\Return $device\_list$
%\EndProcedure
%\Return $received\_data$
%\end{algorithmic}
\end{algorithm}

%\begin{function}[tb]
%\caption{Listen($ch,dwell\_time$)}
%\label{alg:listen}
%\begin{algorithmic}[1]
% \Require $n \geq 0$
% \Ensure $y = x^n$
%\SetKwComment{Comment}{$\triangleright$\ }{}
%\textbf{procedure} LISTEN($ch,dwell\_time$)\\
%\Comment{Receive packets on channel $ch$ for duration $dwell\_time$ and return list of discovered devices}
%\Comment{A channel $ch$ is characterized by a tuple ($freq, bw, proto$), i.e., its frequency $ch.freq$, channel bandwidth $ch.bw$ and protocol $ch.proto$.}
%\Procedure{Listen}{$ch,dwell\_time$}\\
%\State $t \gets 0$ \Comment*[f]{Initialize timer}\\
%\State $device\_list \gets\{\}$ \Comment*[f]{Initialize device list}\\
%\While{$t \le dwell\_time$}{
%\State Listen on channel $ch$ \\
%\State Get packet and extract device address $dev\_addr$  \label{alg:listen:empty}\\
%\Comment{Note that if packet does not contain a device address, $dev\_addr$ is an empty element.} 
%\State $device\_list = device\_list \cup dev\_addr$ \label{alg:listen:add-dev}\\
%\Comment{Add device to list.} 
%{Note that if device is already an element in the list, union operation does not change list.}  
%}
%\EndWhile
%\State \Return $device\_list$
%\EndProcedure
%\Return $received\_data$
%\end{algorithmic}
%\end{function}

The key building block to any of the following scanning algorithms is the function \textbf{Listen()} (Algorithm~\ref{alg:listen}). It takes two input parameters, namely a channel $ch$ (defined by a center frequency, bandwidth and protocol) and a time period $dwell\_time$ after which the procedure terminates listening to channel $ch$. During execution of this procedure, the SDR decodes any packet received on the channel, and extracts address information $dev\_addr$ that identifies a device (line~\ref{alg:listen:empty}). Note that some packets may have no address information, in which case $dev\_addr$ is an empty set. Next, the device address is added to the list of discovered devices $device\_list$ (line~\ref{alg:listen:add-dev}). By definition, if $dev\_addr$  already appears in $device\_list$, then the union operation does not change the contents of the list. Upon the expiration of the channel dwelling time, the procedure returns the list of discovered devices. 

% Algorithm 2: Passive Scan
\begin{algorithm}[tb]
\caption{Passive\_Scan($ch\_list, \newline dwell\_time, $ $scan\_time$)}
\label{alg:passive}
\SetKwComment{Procedure}{}{}
\SetKwComment{Comment}{$\triangleright$\ }{}
%\begin{algorithmic}[1]
%\Procedure{Passive\_Scan}{$ch\_list,dwell\_time,scan\_time$}
%\textbf{procedure} PASSIVE\_SCAN($ch\_list, dwell\_time,$ \pushline\dosemic\nonl $scan\_time$) \\
\Comment{Enumerate devices by repeatedly listening for duration $dwell\_time$ on each channel in $ch\_list$ and stop after $scan\_time$}

\SetKwComment{Comment}{$\triangleright$\ }{}

% \Require $n \geq 0$
% \Ensure $y = x^n$
%\Comment*[h]{Total number of channels}
%\State $C \gets |ch\_list|$ \\

%\Comment*[h]{Set first channel to listen on}
%\State $ch \gets 0$ \\

$t_{start} \gets $time() \Comment*[f]{Store current time}\\
$device\_list \gets \{\}$ \Comment*[f]{Initialize device list}\\
$i \gets 0$\Comment*[f]{Set channel counter to zero}\\
% \While{$t \le scan\_time$}{
\While{\textnormal{time()}$-t_{start} \le scan\_time$}{
\Comment{$ch\_list(i)$ is the $i$-th element in $ch\_list$}

$new\_dev \gets$ \textbf{Listen}{($ch\_list(i), dwell\_time$)} \\
$device\_list = device\_list \cup new\_dev$\\
%\Comment*[h]{Go to next next channel}\\
%\State $ch \gets (ch + 1) \mod C $ \\
$i \gets (i+1)$ mod $|ch\_list|$\\
%\If{$t \geq scan\_time$}{\Comment{Terminate when $scan\_time$ has passed.}\\\textbf{break}} 
}
%\EndWhile

\Return $device\_list$ 
%\EndProcedure
%\While{$discovered\_devices \le DEVICES$}
%\State $received\_data \gets rx(CHANNELS[channel],TIMEOUT)$
%\If{$known\_devices$ is in $received\_data$}
%    \State $discovered\_devices \gets discovered\_devices + 1$
    % \State $N \gets \frac{N}{2}$  \Comment{This is a comment}
%\EndIf
%\If{$channel == length(CHANNELS)$}
%    \State $channel \gets 0$
%\Else
%\State $channel \gets channel + 1$
% \ElsIf{$N$ is odd}
%     \State $y \gets y \times X$
%     \State $N \gets N - 1$
%\EndIf
%\EndWhile
%\end{algorithmic}
\end{algorithm}
%\end{document}

Algorithm~\ref{alg:passive} presents a simple sequential scanning procedure~\textbf{Passive\_Scan} that can be used in conjunction with any IoT protocol. This algorithm represents a baseline against which the performance of more advanced algorithms can be compared. The algorithm invokes the \textbf{Listen} procedure in a round-robin fashion on each channel of a given channel list $ch\_list$, which is provided as an input to the procedure. The total scan time is set by the $scan\_time$ input parameter. Note that generally $scan\_time \gg dwell\_time$, and hence each channel is visited several times during the scan. The algorithm returns the list of discovered devices.

% Algorithm 3, 4: Probe Channels, Active Scan
\begin{algorithm*}[tp]
\caption{Probe\_Channels($ch\_list,dwell\_time$)}
\label{alg:probech} % alg:probech
\SetKwComment{Comment}{$\triangleright$\ }{}
%\textbf{function}$\,$PROBE\_CHANNELS($ch\_list,dwell\_time$) \\
\Comment{Actively probe each channel $ch$ from the channel list $ch\_list$ for duration of $dwell\_time$.}
$active\_channels \gets \{\}$\\
$device\_list \gets \{\}$\\
\For{$ch \in ch\_list$}{ 
Send probe on channel $ch$ \Comment*[f]{Trigger responses}\\
    $new\_dev \gets$ \textbf{Listen}{($ch,dwell\_time$)}\\
    \If{$new\_dev \neq \{\}$}{ %\emptyset$}{
    %\Comment{New device(s) imply an active channel: Store device(s) and channel.}
    $device\_list \gets device\_list \cup new\_dev$ \Comment*[f]{Add found devices}\\
    $active\_channels \gets active\_channels \cup ch$ \Comment*[f]{Add channel to active channel list}\\
    }  
}
\Return $active\_channels, device\_list$
\end{algorithm*}

\begin{algorithm*}[tp]
\caption{Active\_Scan($ch\_list,dwell\_time,scan\_time$)}\label{alg:active}
\SetKwComment{Comment}{$\triangleright$\ }{}
%\begin{algorithmic}[1]
%\Procedure{Active\_Scan}{$ch\_list,dwell\_time,scan\_time$}
%\textbf{function} ACTIVE\_SCAN($ch\_list,dwell\_time,scan\_time$)\\
\Comment{Enumerate devices by first identifying the list of $active\_channels$ in $ch\_list$ and then performing passive scanning only on those $active\_channels$}

% \Require $n \geq 0$
% \Ensure $y = x^n$
%\State $C \gets |ch\_list|$ \Comment{Total number of channels}
$device\_list \gets \{\}$ \Comment*[f]{Initialize list of found devices}\\
%\State $ch \gets 0$ \Comment*[f]{Set first channel to listen on}\\
$active\_channels \gets \{\}$ \Comment*[f]{Initialize list of busy channels}\\
$t_{start} \gets $time() \Comment*[f]{store current time}\\
%\While{$t \le scan\_time$}{
% \For{$ch \in ch\_list$}{ \Comment*[f]{First scan each channel and find if it is busy}\\
% \label{phase1-start}
% \State Send probe on channel $ch$\\
% \State $new\_dev \gets$ \textbf{Listen}{($ch,dwell\_time$)}\\
% \If{$new\_dev \neq \emptyset$}{ \Comment*[f]{New device(s) found implying a busy channel}\\
% \State  $device\_list \gets device\_list \cup new\_dev$ \\
% \State $busy\_channels \gets busy\_channels \cup ch$
% }
% %\EndIf
% }

\Comment{Phase 1: Scan active devices}
 $active\_channels, device\_list \gets \textbf{Probe\_Channels}(ch\_list,dwell\_time)$ \label{alg:active:phase1scan} \label{alg:active:phase1-start}\\% blocking call to this procedure
%\EndFor
\label{alg:active:phase1-end}

\Comment{Phase 2: Passive-scan known active channels for the remaining time}
$t_{scan} \gets scan\_time - (\textnormal{time()}-t_{start})$ \label{alg:active:phase2-start} \Comment*[f]{Compute remaining scanning time}\\
$new\_dev \gets \textbf{Passive\_Scan}(active\_channels, dwell\_time, t_{scan})$  \Comment*[f]{Run passive scan on active channels} \label{alg:active:phase2scan} \label{alg:active:passive}\\
$device\_list \gets device\_list \cup new\_dev$ \Comment*[f]{Add devices found during passive scanning}\\
\label{alg:active:phase2-end}
%}
%\EndWhile
\Return $device\_list$ 
%\EndProcedure
%\end{algorithmic}
\end{algorithm*}

\begin{figure}[t]
    \centering
    \includegraphics[width=\linewidth]{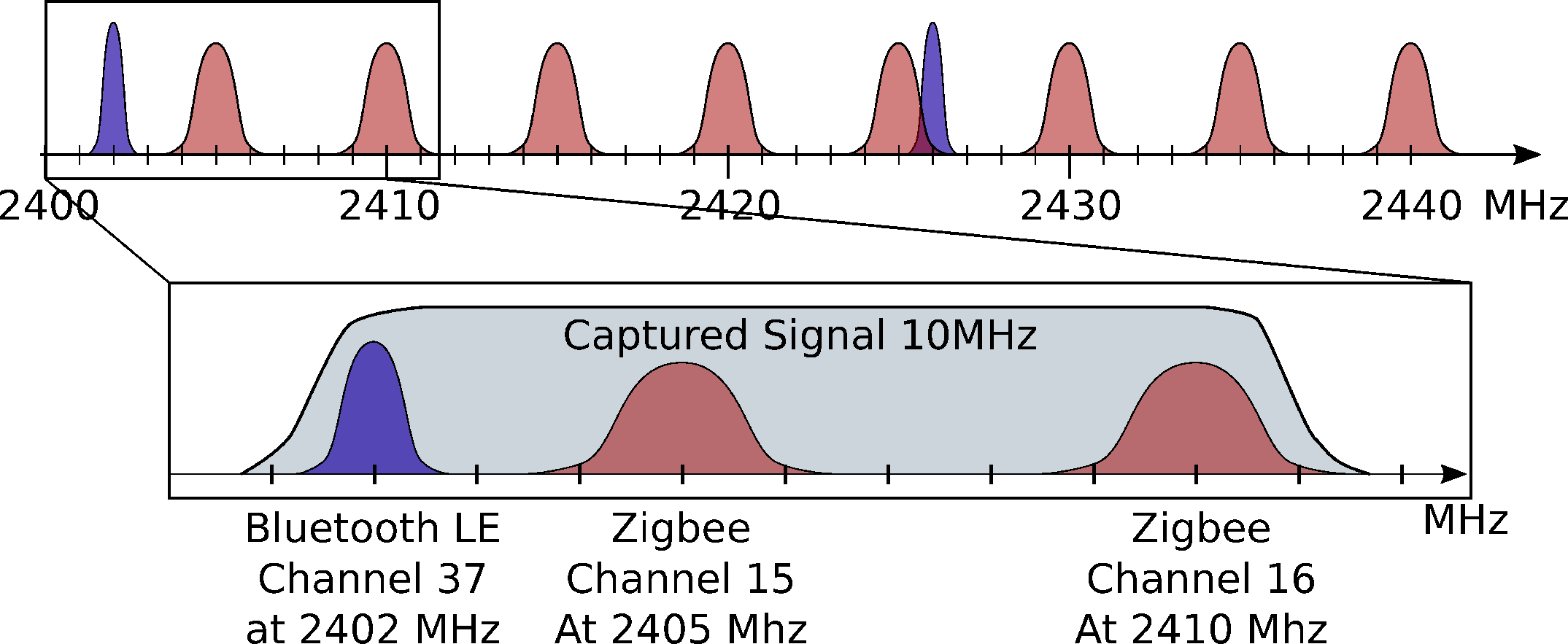}
    \caption{\textbf{Find\_Channels\_In\_Range()} (Algorithm~\ref{alg:chinrange}) starts at the lowest channel in the provided list and returns all channels that are in range of the SDR hardware based on the provided instantaneous bandwidth parameter.}
    \label{fig:illustration-chinrange}
\end{figure}

Sequential passive scanning can be slow, especially if an IoT protocol supports many channels, but only a few channels are used. In order to speed up device discovery, Algorithm~\ref{alg:active}, which we refer to as \textbf{Active\_Scan}, implements a two-phase approach. During the first phase (line~\ref{alg:active:phase1-start}), it invokes a helper function \textbf{Probe\_Channels} (Algorithm~\ref{alg:probech}), which sends a probe packet on each channel~$ch$ in the provided $channel\_list$ and waits for a response. If one or more devices respond, then channel $ch$ is added to the $active\_channels$ list. During the second phase (lines~\ref{alg:active:phase2-start}--\ref{alg:active:phase2-end}), Algorithm~\ref{alg:active} performs passive scanning only on channels appearing in the $active\_channels$ list for the remaining scan time. Algorithm~\ref{alg:active} is especially useful for protocols such as Zigbee, which defines 16 different channels, not all of which may be in use. For Zigbee, {\tt IoT-Scan} implements the probe packet using a \emph{beacon request} frame, to which Zigbee coordinators and routers respond with a \emph{beacon} frame (see also Section~\ref{sec:background}).

\subsection{Multi-channel methods}

The subsequent algorithms expand from single channel scanning to handling multiple channels and multiple protocols, at the same time.

% Algorithm 5: Channels in Range

\begin{algorithm*}[tp]
\caption{Find\_Channels\_In\_Range($ch\_list, bandwidth$)}\label{alg:chinrange}
\SetKwComment{Comment}{$\triangleright$\ }{}

%\textbf{procedure} CHANNELS\_IN\_RANGE($ch\_unscanned, BW$)\\
\Comment{
Identify all channels in $ch\_list$ (ordered by ascending frequency) that can fit in the instantaneous  $bandwidth$, starting from the first element $ch\_list(0).$}
%\State $ch_1 \gets ch\_list(1)$  \Comment*[f]{Select first channel}\\
%\State $ch\_range \gets \{\}$  \Comment*[f]{Init. list of channels in range}\\
% in each inst.bw there can be up to N parallel rx chains/protocols
%\State $start\_freq \gets ch$ \\%ch\_list[0]$\\
%\State $ch\_scan \gets \{ch \in ch\_unscanned \,|\, ch.freq + ch.bw \leq ch\_first.freq+BW$\} \Comment*[h]{Select all ch within BW}\\
%\State Set center frequency $center\_freq$ such that channel $ch_0$ is on the far left of the instantaneous bandwidth $BW$ \label{math1}\\
%\Comment{Add all channels that fit in the current instantaneous bandwidth $BW$ to the list $ch\_range$}
$ch\_range\!\gets\!\{\!$all\,$ch$\,in\,$ch\_list$\,such\,that\,$(ch._{freq}+ch._{bw}/2)-(ch\_list(0)._{freq}-ch\_list(0)._{bw}/2) \leq bandwidth\!$\}\label{alg:chinrange:condition}\\
%\State $center\_freq \gets ch\_first.freq - ch\_first.bw/2 + BW/2$ \Comment*[f]{set first channel on far left of BW} \label{math1} \\
%\State $offset\_list \gets offset\_list \cup (BW/2 - ch.bw/2) $ \Comment*[f]{frequency offset of first channel} \label{math2} \\
%\State $offset\_list \gets \{center\_freq - ch.freq \quad\forall\quad ch \in ch\_range\}$ \Comment*[f]{calculate frequency offsets} \label{math2} \\
\Return $ch\_range$%center\_freq,offset\_list, 
\end{algorithm*} 

Prior to discussing multi-channel and multi-protocol scanning algorithms, we need a method of grouping channels within the range of the instantaneous bandwidth of the SDR.
%in order to scan them simultaneously is required.
\textbf{Find\_Channels\_In\_Range()} (Algorithm~\ref{alg:chinrange}) identifies all channels in an input channel list (ordered by ascending frequency) by selecting all channels that fit the instantaneous bandwidth under consideration of their respective center frequencies and channel bandwidths, see Fig.~\ref{fig:illustration-chinrange}. For a channel to be considered in range, the entire bandwidth of the signal must be contained in the captured instantaneous bandwidth of the SDR (see line~\ref{alg:chinrange:condition} of Algorithm~\ref{alg:chinrange}).

In practice, the center frequency of each group of channels is set such that the first channel from the channel list (i.e., the one with the lowest frequency) is at the far left end of the instantaneous bandwidth.
%By doing so, the algorithm tries to include subsequent channels from the list in its current bandwidth.
If none of the other channels' bandwidths overlap with the current instantaneous bandwidth, the function will return the first element of the input channel list, i.e., it will default to single-channel selection.

% Algorithm 6: Listen in Parallel
\begin{algorithm*}[tp]
\caption{Listen\_In\_Parallel($ch\_range, dwell\_time$)}\label{alg:listenparallel}
%\begin{algorithmic}%[1]
\SetKwComment{Comment}{$\triangleright$\ }{}

\SetKwBlock{DoParallel}{do in parallel}{end parallel}
%\textbf{procedure} LISTEN\_IN\_PARALLEL($ch\_range, dwell\_time$)\\
\Comment{
%Each element of channel list $ch\_list$ is a tuple ($freq, bw, proto$). Hence, channel $ch$ in channel list is defined by its frequency $ch.freq$, channel bandwidth $ch.bw$ and protocol $ch.proto$.
Scan in parallel all channels in $ch\_range$ for a duration $dwell\_time$. Note that all channels are assumed to be within the instantaneous bandwidth of the SDR, e.g. as produced by Find\_Channels\_In\_Range()}

\DoParallel{
    %\For{$ch \in ch\_range$}{
    %     \State $new\_dev \gets$ \textbf{Listen}{($center\_freq, offset, proto, dwell\_time$)}\\
    %     \State $device\_list \gets device\_list \cup new\_dev$
    % }
    $new\_dev \gets \{$\textbf{Listen}$(ch, dwell\_time)$ for all $ch$ in $ch\_range$\}\\
    $device\_list \gets device\_list \,\cup new\_dev$ \\
    %}
}
\Return $device\_list$ 
\end{algorithm*}

We further define a helper function \textbf{Listen\_In\_Parallel()} (Algorithm~\ref{alg:listenparallel}) which simultaneously listens to multiple channels by calling \textbf{Listen()} (Algorithm~\ref{alg:listen}) on all provided channels. Note that implementing this algorithm requires extracting multiple signal streams by frequency-shifting, filtering, and down-conversion or resampling the incoming signal relative to its center frequency and the parameters of the channel. This procedure is called \emph{channelization}. The implementation aspects of this procedure are described in Section~\ref{sec:gnuradio-processing}.

% Algorithm 7: Multiprotocol Scan

\begin{algorithm*}[tp]
\caption{Multiprotocol\_Scan($ch\_list, dwell\_time, scan\_time, bandwidth$)}\label{alg:multiproto}
\SetKwComment{Comment}{$\triangleright$\ }{}
\Comment{This algorithm enumerates devices by scanning as many channels as can fit in the instantaneous bandwidth of $bandwidth$ for a duration $dwell\_time$ in each iteration.}
\SetKwBlock{Loop}{loop}{end loop}
$ch\_unscanned \gets ch\_list$ \Comment*[f]{All channels in the list are unscanned}\\
$ch\_groups \gets \{\}$ \Comment*[f]{Initialize list of channel groups}\\
\While{$ch\_unscanned \neq \{\}$}{
        \Comment{Find channels that can be scanned simultaneously as they fit the instantaneous $bandwidth$.}
        $ch\_range \gets$ \textbf{Find\_Channels\_In\_Range}(ch\_unscanned, bandwidth)\\%removed center\_freq,offset\_list,
        \Comment*[h]{Scan channels that fit in instantaneous bandwidth BW around center freq.}
        $ch\_groups \gets ch\_groups \cup \{ ch\_range \}$ \Comment*[f]{Add this group to the list of channel groups}\\
        $ch\_unscanned \gets ch\_unscanned \setminus ch\_range$ \Comment*[f]{Remove channels from unscanned list}
}
$t_{start} \gets $time() \Comment*[f]{Store current time}\\
$device\_list \gets \{\}$ \Comment*[f]{Initialize list of found devices}\\ %
$i \gets 0$ \Comment*[f]{Set channel counter to zero}\\
\While{\textnormal{time()}$-t_{start} \leq scan\_time$}{
        \Comment{Scan all channels of the i'th channel group in parallel}
        $new\_dev \gets $ \textbf{Listen\_In\_Parallel}(ch\_group(i), dwell\_time)\\
        \Comment{Remove scanned channels from unscanned channel list}
        $device\_list \gets device\_list \,\cup new\_dev$
        $i \gets (i+1)$ mod $|ch\_groups|$
}
\Return $device\_list$ 
\end{algorithm*}

%\begin{algorithm*}
%\caption{Multiprotocol\_Scan($ch\_list, dwell\_time, scan\_time, bandwidth$)}\label{alg:multiproto}
%\SetKwComment{Comment}{$\triangleright$\ }{}
%\Comment{This algorithm enumerates devices by scanning as many channels from the channel list in parallel as can fit in the instantaneous bandwidth of $bandwidth$ in duration $dwell\_time$ on each channel hop.}\\

%\SetKwBlock{Loop}{loop}{end loop}

%\State $device\_list \gets \{\}$ \Comment*[f]{Initialize list of found devices}\\ \\%

%\State $t_{start} \gets $time() \Comment*[f]{store current time}\\
%\Loop{
%    \State $ch\_unscanned \gets ch\_list$ \Comment*[f]{All channels in the list are unscanned}\\
%    \While{$|ch\_unscanned| > 0$}{
%        \Comment{Select channels that can be scanned simultaneously as they fit the instantaneous $bandwidth$.} \\
%        \State $ch\_range$ \gets \textbf{Find\_Channels\_In\_Range}(ch\_unscanned, bandwidth)\\%removed center\_freq,offset\_list,
%        \Comment*[h]{scan channels that fit in instantaneous bandwidth BW around center freq.} \\
%        \SetKwBlock{DoParallel}{do in parallel}{end}
%        \State $new\_device$ \gets \textbf{Listen\_In\_Parallel}$(ch\_range, dwell\_time)$ \\
%        \State $device\_list \gets device\_list \,\cup new\_device$
%        \Comment*[h]{remove scanned channels from unscanned channel list}\\
%        \State $ch\_unscanned \gets ch\_unscanned \setminus ch\_range$ \\
%        \If{$t \geq scan\_time$}{
%            \textbf{break\Comment*[f]{Terminate when $scan\_time$ has passed.}\\}
%        } 
%    }
%}
%\State \Return $device\_list$ 
%
%\end{algorithm*}

Algorithm~\ref{alg:multiproto} describes a parallel multi-protocol scan that can be used with any number of IoT protocols. Based on a list of channels to consider (ordered by ascending frequencies), the algorithm starts at the lowest frequency and determines all channels within range of the first channel by calling \textbf{Find\_Channels\_In\_Range()} (Algorithm~\ref{alg:chinrange}). It subsequently listens to those channels by invoking \textbf{Listen\_In\_Parallel()} (Algorithm~\ref{alg:listenparallel}). Note that if only one channel is in range at given step of the while loop (line 12), then the algorithm's behavior becomes identical to \textbf{Passive\_Scan()} (Algorithm~\ref{alg:passive}).

Each such channel hop is scanned for the defined channel $dwell\_time$. Once all unscanned channels are exhausted, the algorithm restarts from the lowest channel until the desired $scan\_time$ has elapsed. Note that the total $scan\_time$ is typically much greater than the channel $dwell\_time$. Depending on the frequency allocation of the protocols involved, the multi-protocol scan algorithm can significantly speed up IoT device discovery process by receiving multiple protocols simultaneously, as demonstrated in Section~\ref{sec:results}.

% Algorithm 8: Active Multiprotocol Scan
\begin{algorithm*}[tp]
\caption{Active\_Multiprotocol\_Scan($ch\_list, ch\_probe\_list, dwell\_time, scan\_time, bandwidth$)}\label{alg:active-multiproto}
%\begin{algorithmic}%[1]
\SetKwComment{Comment}{$\triangleright$\ }{}
%\textbf{procedure} ACTIVE\_MULTIPROTO\_SCAN($ch\_list, dwell\_time, scan\_time, bandwidth$)\\
\Comment{Enumerate devices by first identifying list of $busy\_channels$ from $ch\_probe\_list$ and then performing multi-protocol scanning only on those active channels and on other channels provided in $ch\_list$.}
% \Require $n \geq 0$
% \Ensure $y = x^n$
%\State $C \gets |ch\_list|$ \Comment{Total number of channels}
$device\_list \gets \{\}$ \Comment*[f]{Initialize list of found devices}\\
%\State $ch \gets 0$ \Comment*[f]{Set first channel to listen on}\\
$active\_channels = \{\}$ \Comment*[f]{Initialize list of busy channels}\\
$t_{start} \gets $ time() \Comment*[f]{Store current time}\\

\Comment{Probe all channels in $ch\_probe\_list$ to identify active channels}
$active\_channels, device\_list \gets $ \textbf{Probe\_Channels}($ch\_probe\_list, dwell\_time$)\\

\label{alg:active-multiproto:phase1-end}
%Create channel list $ch\_list$ based on $busy\_channels$ and BLE advertising channels $ble\_channels$
$t_{scan} \gets scan\_time - (\textnormal{time()}-t_{start})$ \Comment *[f]{Compute remaining scanning time} \label{alg:active-multiproto:phase2-start}\\
\Comment{Merge successfully probed channels with regular channels and sort by ascending frequency}
$active\_channels \gets \textnormal{sort}(active\_channels \cup ch\_list)$ \\
\Comment{Run passive scan on all active channels for the remaining time $t_{scan}$}
$new\_dev \gets$ \textbf{Multiprotocol\_Scan}{($active\_channels,  dwell\_time, t_{scan}, bandwidth$)}  \\
$device\_list \gets device\_list \cup new\_dev$ \Comment*[f]{Add devices found during passive scanning}\\
\label{alg:active-multiproto:phase2-end}
%}
%\EndWhile
\Return $device\_list$ 
\end{algorithm*}

Finally, \textbf{Active\_Multiprotocol\_Scan()} (Algorithm~\ref{alg:active-multiproto}) is a combination of the aforementioned active scanning  and multi-protocol scanning capabilities. It is useful for
%scanning multiple protocols some of whose channels fit in the instantaneous bandwidth of an SDR. We demonstrate this algorithm 
scanning multiple protocols, some actively and some passively (such as a combination of Zigbee and BLE). %The listing of this algorithm appears in Appendix~\ref{sec:active_multi_proto}.
Note that Algorithm~\ref{alg:active-multiproto} receives two lists of channels: $ch\_probe\_list$ and $ch\_list$. Active channels (e.g., Zigbee channels) are only sought among channels in the $ch\_probe\_list$. This step is skipped for channels (e.g., BLE channels) in the $ch\_list$.
\section{Performance Metrics and Analysis}
\label{sec:problem}

In this section, we define metrics to benchmark the various algorithms. We further formalize IoT device discovery as a variation of the non-uniform (weighted) coupon collector problem~\cite{Flajolet1992Coupon,Anceaume2015}. Under appropriate statistical assumptions, the coupon collection time can be computed numerically and serve as a baseline against which the performance of the algorithms can be compared.
%\subsection{Objective}

%Objective of this work is to discover as many IoT devices as possible in the least amount of time. Mainly the focus is on the three popular protocols in the 2.4GHz band: Zigbee, BLE, and LoRa.

\subsection{Metrics}
Our main metric is the discovery time of IoT devices, which we aim to minimize. Assume there are $N$ devices in total,
%labelled  $i=1,2\ldots N$,
with corresponding discovery times  $T_{1},T_{2},\ldots ,T_{N}$. We are interested in characterizing the \emph{order statistics} of these random variables, i.e., the time elapsing till one device is discovered, which is denoted $X_{1:N}$, then till two devices are discovered, which is denoted $X_{2:N}$, and so on, till all devices are discovered, which is denoted
$X_{N:N}$. We thus have
\begin{eqnarray}
%\begin{gather*}
    X_{1:N} &= &\min(T_{1},T_{2},\dots,T_{N}), \\
    X_{2:N} & = &\min(\{T_{1},T_{2}, \dots, T_N\} \setminus  X_{1:N}), \\
    \dots \notag \\
    X_{N:N} & = &\max(T_{1},T_{2},\dots,T_{N}).
    %X_{\text{min}} &=& \text{min}(X_1,\ldots,X_n) \sim  %\text{exp}(\sum_{i=1}^{n}\lambda_i) \\
    %\text{and }\mathbb{E}[X_{\text{min}}] &=& %\frac{1}{\sum_{i=1}^{n}\lambda_i}.
%\end{gather*}
\end{eqnarray}

%Where a random variable $X_{\text{X:N}}$ is the time to discover X-th out of N devices (whichever it may happen to be). A measurement estimate of the arrival time each of X-th device is $E[X_{\text{X:N}}]$. To visualize expectations for all of the order statistics we plot them versus discovered device number, for each scanning algorithm. 
In our experiments, we estimate the expectation of the $n$-th order statistics $E[X_{n:N}]$, for $n = 1, 2, \dots, N$. To obtain these estimates, we run each scanning algorithm $M$ times and denote by $x_{n:N}^{(m)}$ the time till $n$ devices are discovered at the $m$-th iteration, where $m=1,2,\dots,M$. We then compute the \emph{sample mean} for the $n$-th order statistics as follows:
\begin{equation}
    \bar{x}_{n:N}=\frac{\sum_{m=1}^{M} x_{n:N}^{(m)}}{M}. \label{eq:sample_mean}
\end{equation}

We also provide $(1-\alpha)100 \%$ \emph{confidence intervals} for our estimates
\begin{equation}
   [\bar{x}_{n:N}-e_{n:N},
\bar{x}_{n:N}+e_{n:N}],  \label{eq:conf_int}
\end{equation}
based on computing the sample standard deviation $s_{n:N}$ and the confidence interval parameter $e_{n:N}$ as follows:
% $\Xbar$
% $\bar{X}$
\begin{eqnarray}
s_{n:N} & = & \sqrt{\frac{1}{M-1}\sum_{m=1}^{M} (x_{n:N}^{(m)}-\bar{x}_{n:N})^2},\\
e_{n:N} & = & t_{\alpha/2,M-1} \times \frac{s_{n:N}}{\sqrt{M}},
%= 2.262 \times \frac{s_{n:N}}{\sqrt{M}}.
\end{eqnarray}
with $t_{\alpha/2,M-1}$ denoting the $1 - \alpha/2$
quantile of the $t$-distribution with $M - 1$ degrees of freedom~\cite{schmeiser1982batch}. In our experiments, described in Section~\ref{sec:exp}, we run $M=10$ independent iterations for each algorithm and consider 95\% confidence intervals (i.e., $\alpha=0.05$), hence  $t_{\alpha/2,M-1}=2.262$~\cite[Table~1]{schmeiser1982batch}. 
% \subsubsection{Precision}
% Since not all devices were discoverable we define a precision metric as follows:
% \begin{equation}
%     precision = \frac{discovered devices}{total devices},
% \end{equation}
%For more on devices that are hard to discover, see Section~\ref{sec:limitations} on limitations.
% ^^ this table belongs to section 4 %
%
% \subsubsection{Speed}
% % active (20-40\% active really) vs passive zigbee scan
% % Ble vs zigbee vs lora
% Speed of device discovery generally decays exponentially during the scan time. %
% \begin{equation}
%     discovery speed = \frac{devices(t_2)-devices(t_1)}{t_2-t_1},
% \end{equation}
%
% \subsubsection{Time to identify 100\% of devices}
% % main experiment: scan until n device are discovered for protocol x. Only parameter: switching time (for multi-channel scan)
% This metric coincides with the discovery time of the last device on the device discovery versus time plots for different scan algorithms in Section \ref{sec:exp}
%
%
%
% \subsubsection{Number of discovered devices during scan time T}
%
% \color{red}TODO: Unclear if this belongs to metrics or is a list of things that should be somewhere else in the paper... Stefan please check
%
% explain why multi channel works better on lora na d not zigbee and ble
%
% run passive scan for zigbee and ble (and lora once fixed)
%
%
% receive all lora flavors/parameters in parallel (not just multi-channel)

% decide lora scan parameters based on real worlds

% send packet every minute

% which channels? random channel use

% how long of a packet?\color{black}

% Traffic Analysis
\subsection{Theoretical Model}
\label{sec:traffic-analysis}

We next propose a theoretical model to estimate the expectations of order statistics of the discovery time, under appropriate statistical assumption. The analysis further assumes an idealized channel environment where no packet loss occurs (in practice such losses could occur due to imperfect receiver implementation or interference). In Section~\ref{sec:results}, we show that the performance of the scanning algorithms approaches that predicted by the theoretical model, which demonstrates the efficiency of the algorithms.

\subsubsection{Statistical assumptions}

% LoRa 2.4GHz module LAMBDA80C-24S & P9OLAMBDA & - & -\\\hline distributed, i.e., $X_i \sim \text{exp}(\lambda_i)$, and result in mean inter-arrival times as well as a standard deviations of $\mu_i=\sigma_i=1/\lambda_i$.

To model device enumeration, we need statistics of the inter-arrival times of packets generated by each device. For the sake of analytical tractability, we assume that devices transmit in a memoryless fashion, i.e., the inter-arrival times of their packets follow an exponential distribution. Note that the mean and standard deviation of an exponential random variable are equal. 
%A renewal process is a counting process where the time duration between events are i.i.d., but not necessarily exponentially distributed. However, it is a known property that under appropriate scaling, the sum of renewal processes converges to a Poisson process, cf. ~\cite[Theorem~9.1, p.~223]{Karlin75}. We 
Hence, we expect that this model can provide a reasonable approximation, 
%of real-world device detection if either individual devices transmit in a memoryless fashion, or the aggregate traffic follows a Poisson process, in which case 
if for each device $i$, its mean inter-arrival time $\mu_i$ and standard deviation of inter-arrival times $\sigma_i$ are roughly equal.   

To check this assumption, we collected statistics of the inter-arrival times of packets of the Bluetooth and Zigbee devices listed
in Table~\ref{tab:testeddevices} below. Specifically, for each device $i$, we measure the times of packet arrivals with $K+1$ timestamps.  We then calculate the $K$ inter-arrival times $\tau_{i,k}=t_{i,k+1}-t_{i,k}$, where $k=1,2,\dots,K$. Based on this data, we obtain estimates of the expectation for each device $i$
\begin{equation}
    \mu_i = \frac{1}{K} \sum_{k=1}^{K} \tau_{i,k}, \label{eq:device-mu}
\end{equation}
as well as the standard deviation
\begin{equation}
%    \mu_i = \frac{1}{K} \sum_{k=1}^{K} \tau_{i,k}, \label{eq:device-mu} \\
    \sigma_i = \sqrt{\frac{1}{K} \sum_{k=1}^{K}(\tau_{i,k}-\mu_i)^2 }.  \label{eq:device-sigma}
\end{equation}
Table~\ref{tab:testeddevices} indicates that indeed for all tested BLE devices $\mu_i \approx \sigma_i$, while this also holds for many, though not all Zigbee devices. 
%Note that for each device $i$, $\mu_i$ and $\sigma_i$ were estimated by collecting timestamps of packet arrivals over the course of two hours. 
%Hence, one should expect the theoretical model to be somewhat less accurate in that case, as observed in Section~\ref{sec:ana_results}.

\begin{comment}
\subsubsection{First and last detected device.}

To model detection of the first device, let
\begin{eqnarray}
    X_{\text{min}} &=& \text{min}(X_1,\ldots,X_n) \sim  \text{exp}(\sum_{i=1}^{n}\lambda_i) \\
    \text{and }\mathbb{E}[X_{\text{min}}] &=& \frac{1}{\sum_{i=1}^{n}\lambda_i}.
\end{eqnarray}

Further, the last detected device can be modeled by 
\begin{eqnarray}
    X_{max} &=& \text{max}(X1,\ldots,Xn) \\
    Pr(X_{max} \leq x) &\stackrel{\mathclap{\footnotesize\mbox{by indep.}}}{=}& \prod_{i=1}^{n}Pr(X_i\leq x) = \prod_{i=1}^{n} (1-e^{\lambda_i x}).
\end{eqnarray}
Then,
\begin{eqnarray}
    \mathbb{E}[X_{\text{max}}] &=& \int_{0}^{\infty} Pr(X_{\text{max}} > x) dx \nonumber\\
    &=& \int_{0}^{\infty} 1 - Pr(X_{\text{max}} \leq x) dx \nonumber\\
    &=& \int_{0}^{\infty} 1 - \prod_{i=1}^{n} (1-e^{-\lambda_i x}) dx \\
    \text{with }X_{\text{max}} &=& \text{max}(X_1,\ldots,X_n).
\end{eqnarray}

While the latter is hard to evaluate, the minimum expectation is easily computed based on the assumption of exponentially-distributed inter-arrival times, i.e. $\lambda_i=1/\mu_i$ for the devices under test.
\end{comment}

\subsubsection{Analysis of order statistics} Enumerating devices shares similarities with the non-uniform coupon collector's problem~\cite{Flajolet1992Coupon},
%~\cite[p.~216ff.]{Flajolet1992Coupon}, 
albeit with certain modifications.

The coupon collector's problem assumes a probability distribution in which each draw results in a coupon (i.e., a discovered device). This cannot be applied directly to a scenario in which devices' transmission characteristics may result in \emph{null coupons}, i.e., a scan iteration in which no new device is discovered. 
Anceaume et al.~\cite{Anceaume2015} provide a method of calculating the expectation of the non-uniform coupon collector problem which accounts for a null coupon. 
Define the probability vector $\boldsymbol{p}$ in which $p_0$ is the probability of no device transmitting, and $p_i$ is the probability of device~$i$ transmitting, $i=1,2,\dots,N$.
The expectation for the $n$-th order statistics $X_{n,N}$ (i.e., the time to to discover $n$ out of $N$ devices) is then given by
%expressed as~\cite[p.~411]{Anceaume2015}:
\begin{eqnarray}
    %\mathbb{E}[X_{n:N}(\boldsymbol{p})] &=& \sum_{i=0}^{n-1}(-1)^{n-1-i}{N-i-1 \choose N-n} \sum_{J \in S_{i,N}} \frac{1}{1-(p_0+P_J)}. \label{eqn:exp_tcn} \\
    \mathbb{E}[X_{n:N}(\boldsymbol{p})] \!& = &\! \sum_{h=0}^{n-1} R_{N,n,h} \! \sum_{J \in S_{h,N}}\!\frac{1}{1-p_0-P_J} \label{eqn:exp_tcn} \\
    \text{with} \nonumber \\
    R_{N,n,h} & = & (-1)^{n-1-h} {N-h-1 \choose N-n}.
\end{eqnarray}

Here, $S_{h,N}$ denotes all $N \choose h$ subsets containing exactly $h$ devices.
%probabilities such that $|J|=i$. For example,
%\begin{eqnarray}
%    S_{1,N} &=& \{(p_1),(p_2),\ldots,(p_N)\} \nonumber \\
%    S_{2,N} &=& \{(p_j,p_k) \text{ such that } j,k \in 1\ldots %N,j \neq k\}  \nonumber \\
%    S_{3,N} &=& \{(p_j,p_k,p_l) \text{ s.t. } j,k,l \in 1\ldots %N,j \neq k \neq l\},  \nonumber
%\end{eqnarray} and so forth.
Denote by $J$ any subset of $S_{h,N}$ that contains exactly $h$ devices. Then, $P_J=\sum_{j \in J}p_j$ is the summation of the transmission  probabilities of all devices belonging to $J$.  Note that the second summation term in Eq.~(\ref{eqn:exp_tcn}) works out to a summation over all possible subsets $J$ of cardinality $h$.
%of the term $\frac{1}{1-(p_0+\sum_{j \in J}p_j)}$.
%We refer to Anceaume et al.~\cite[p.~411f.]{Anceaume2015} for more detailed discussions of these terms.

Assuming all $N$ devices send packets in an i.i.d. memoryless fashion as discussed above, the device traffic can be modeled as $N$ independent Poisson processes with rate $\lambda_i = 1/\mu_i$. The combined influx of packets from all the devices then follows a Poisson process with rate $\lambda = \sum_{i=1}^{N} \lambda_i$.
%\begin{equation}
%    Poisson(\sum_{i=0}^{n} \lambda_i)= Poisson(\lambda). %\label{eq:poisson-lambda}
%\end{equation}

By selecting a small interval $\Delta t$ such that either zero or one packet arrives during any  interval $\Delta t$, we can use Eq.~(\ref{eqn:exp_tcn}) to compute the expectation of the order statistics of the discovery time of devices. Let $Z$ be a Poisson random variable with mean $\lambda \Delta t$ that counts the number of packets arriving from all devices during a time interval $\Delta t$. We have
\begin{eqnarray}
    \Pr(Z=0) &=& 
    %\frac{(\Delta t\lambda)^{0} e^{-\Delta t\lambda}}{0!}=\Delta t 
    e^{-\lambda \Delta t}, \label{eq:prz0}\\
    \Pr(Z=1) &=& 
    %\frac{\Delta t\lambda e^{-\Delta t\lambda}}{1!}=
    (\lambda \Delta t) e^{-\lambda \Delta t}, \label{eq:prz1} \\
    \Pr(Z \geq 2) &=& 1-Pr(Z=0)-Pr(Z=1). \label{eq:przlt1}
\end{eqnarray}
In order to determine a suitable $\Delta t$, we select it such that $Pr(Z\!\geq\!2 )$ becomes negligible, as discussed in Section~\ref{sec:results}. 

If all devices transmit on one channel that is continuously monitored, the probability $p_i$ that device $i$ transmits during an interval $\Delta t$ is then
\begin{equation}
p_i = (\lambda_i \Delta t) e^{-\lambda_i \Delta t} \approx \lambda_i \Delta t.
    %p_{i,\text{single}} &=& Pr(X_i < \Delta t) = 1 - e^{-\lambda_i \Delta t}, 
    \label{eq:proba-single}
    %\text{and } \quad p_0 &=& 1-\sum_{i=1}^{n}p_{i,\text{single}}\,\forall i=1\ldots n.
\end{equation}

Note that if all devices are randomly distributed on any of $C$ available channels, a randomly channel-hopping radio scanner would receive a transmission from device $i$ with  probability $p_i/C$. This can also be used as an approximation when the scanner visits channels in a round-robin rather than in a random fashion.
%hence 
%\begin{equation}
%p_i = (\lambda_i \Delta t) e^{-\lambda_i \Delta t}/C \approx \lambda_i \Delta t/C.
    %p_{i,\text{single}} &=& Pr(X_i < \Delta t) = 1 - e^{-\lambda_i \Delta t}, \label{eq:proba-single}\\
    \label{eq:proba-prx0}

\section{Protocol Device Enumeration}
\label{sec:background}

In the previous sections, the concepts of ``listening to a channel'' and ``extracting device addresses''  were presented in a generic way (see Algorithm~\ref{alg:listen}). We now discuss these aspects in detail for all the IoT protocols implemented in {\tt IoT-Scan}. This section gives an overview of implementation-specific considerations for protocols covered in this work.

%In the previous sections, the concepts of ``listening to a channel'' and ``extracting device addresses''  were presented in a generic way (see Algorithm~\ref{alg:listen}). We now discuss these aspects in detail for all the IoT protocols implemented in {\tt IoT-Scan}. 

%This appendix provides implementation-specific considerations  for each of the IoT protocols covered in this work. In particular, we detail how {\tt IoT-Scan} extracts device addresses in each case.
%This lower-level information is necessary for the practical implementation of the scanning algorithms of all the protocols on a common SDR platform.

%In the following sections, we will briefly give an overview of protocols covered in this work, and discuss technical details for device enumeration, and other implementation considerations.

%In this section, we discuss the methods used for enumerating devices under the various IoT protocols. Specifically, we discuss channels, packet structure, and addressing fields in particular and how to enumerate devices using those addressing fields.
%This section provides an overview of the protocols we are addressing in this work.
%\color{red}
%\begin{verbatim}
%TO DO: Content
%    IoT protocols
%    - description
%    - channels
%    - relevant packet contents
%    ...etc.
%\end{verbatim}
%\color{black}
\subsection{Zigbee}
\label{sec:proto:zigbee}
Zigbee is a network and application layer protocol which uses the IEEE~802.15.4 physical layer specification~\cite{ieee802.15.4}. It is widely used in home and commercial building automation applications such as lighting, climate, and access control~\cite{ZigbeeWhatIs}. Zigbee operates on 16 channels on the 2.4GHz ISM band. Each channel is 2\,MHz wide and centered at $f_c = 2405 + 5(k-11)$\,MHz for channels $k=11...26$~\cite[p.~387]{ieee802.15.4}. 

Zigbee defines three types of devices: \emph{coordinators}, \emph{routers}, and \emph{end-devices}, each of which behave differently on the network. End-devices do not route traffic, and are typically mobile and battery-powered, i.e. energy-constrained. As a result, end devices are frequently sleeping, i.e., remain inactive in order to save power. Routers route traffic, receive and store messages for their children (i.e., end devices that they route traffic from and to), and communicate with new nodes requesting to join the network. Therefore, routers cannot sleep and are typically mains-powered devices. A Zigbee coordinator is a special router which, in addition to all of the router capabilities, also forms a network. Before creating a network, Zigbee coordinators scan available channels to select a good, i.e. low interference, channel for the network.
% This is done by performing an energy detection scan over all channels, and selecting the channels with the least RF activity. Then, the coordinator performs an active scan on those found good channels by sending (broadcast) beacon requests and then listening for any response beacons \cite{digi}. After network creation the coordinator allows other routers and end-devices to join (i.e., connect to) the network. When routers and end-devices try to join the network they first scan one or more channels in search of a network. 

\begin{figure}[t]
    \centering
    \includegraphics[width=\linewidth]{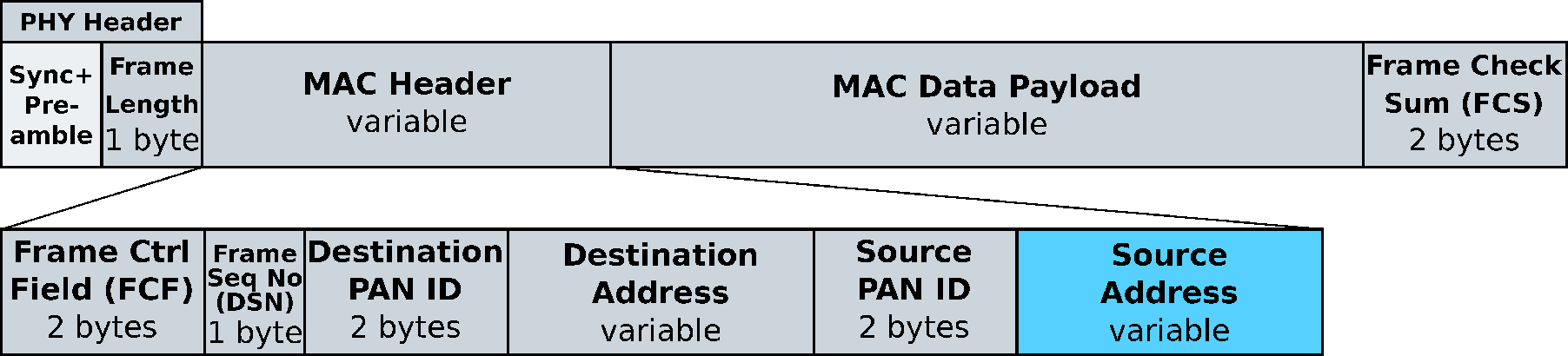}
    \caption{Zigbee medium access control (MAC) layer frame structure \cite{ZigbeeWhatIs}.}
    %\color{red}TODO: Figure does not reflect different address/frame types mentioned in text\color{black}
    \label{fig:zigbee-frame-structure}
\end{figure} 

\textbf{Address Information.} {\tt IoT-Scan} enumerates Zigbee devices by both their \emph{short} (16~bits) and \emph{extended} (64~bits) address (whichever of those address types is present in a given packet), ensuring no device is counted double despite these two address formats. While short addresses are unique within a network, extended addresses are typically assigned by the manufacturer in a globally unique way. Some Zigbee packet types will use short and some will use a long address. Both addresses can be parsed from the ``Source Address'' variable-length field in Figure~\ref{fig:zigbee-frame-structure}.
%These addresses can appear in different parts of the frame, such as in the network layer frame and IEEE802.15.4 MAC layer (Fig.~\ref{fig:zigbee-frame-structure}).
%The position of these addresses is shown in Figure~\ref{fig:zigbee-frame-structure}.
Note that the PAN ID (personal area network identifier), shown in the same figure, is a network identifier. While we do not use it in our scanning, it could be useful to differentiate between two devices with the same short address but on different networks.

\begin{figure}[tb]
    \centering
    \includegraphics[width=\linewidth]{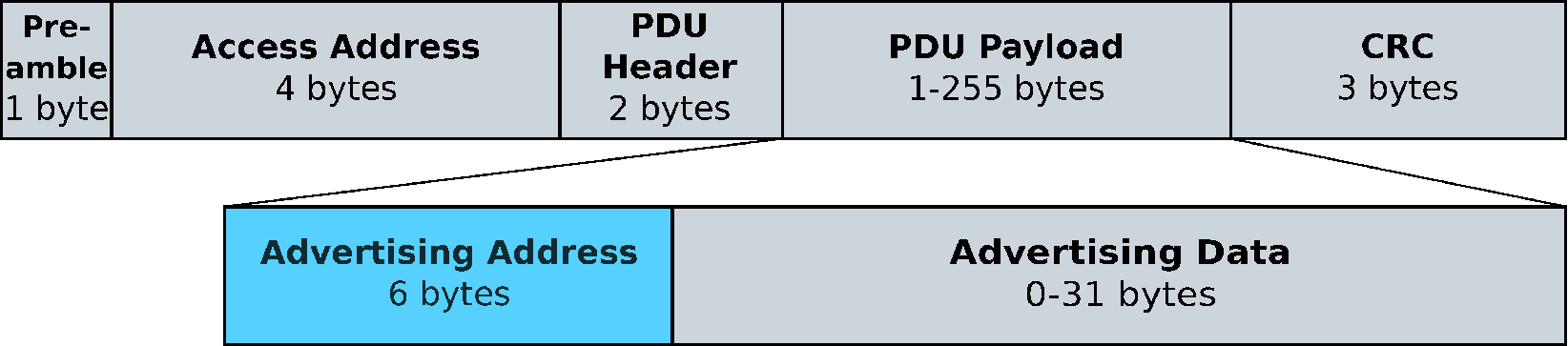}
    \caption{The BLE advertising packet format (top) and the Advertising Channel PDU Header (bottom) for common advertising messages (PDU Types ADV\_IND, ADV\_NONCONN\_IND, ADV\_SCAN\_IND) \cite[p.~2562,~2567]{BluetoothCoreSpec5}.}
    \label{fig:ble-frame-structure}
\end{figure} 
%For instance, data and MAC command frames typically have both 16bit short and extended 64bit address. Short addresses appear both in MAC (a.k.a. 802.15.4 or WPAN) layer and MAC data payload (a.k.a. Zigbee network layer) as shown in Figure \ref{fig:zigbee-frame-structure}. While extended addresses appear only in Zigbee network layer. On the other hand, beacon requests
%(frame type = command, command id = 7)
%contain only broadcast address 0xffff and no exact source address that could be used for enumeration. However, beacon frame types do contain useful short address of the beacon sender.

% Enumerating Zigbee devices requires considering different field types depending on the frame type in question (beacon, data, ACK, MAC command) to collect address information, which may be in short (16~bits) or extended (64~bits) address form. In the case of acknowledgment (ACK) frames, there is no address at all. Frame type and addressing mode is specified in the frame control field (FCF), see Fig.~\ref{fig:zigbee-frame-structure}.

\textbf{Implementation.} {\tt IoT-Scan} implements both passive and active scans for Zigbee.
%(Algorithm~\ref{alg:passive}) and active (Algorithm~\ref{alg:active}) multi-channel Zigbee scans using an SDR. 
A passive scan listens on each channel for a certain amount of time (i.e., the \emph{channel dwell time} in Algorithm~\ref{alg:listen}) repeatedly until the total scan time expires. 
%Note that  since passively scanning relies on the transmission behavior of devices out of the scanner's control, passive device enumeration may take an arbitrary amount of time. % to the probabilistic nature of traffic a passive scanner is able to receive in a limited amount of time, usually a passive scanner requires multiple rounds to discover a certain number of devices,
%while an active scanner requires a single round to find all active channels, and the coordinator and router devices on these channels. 
With active scanning, channels with network activity are discovered by sending beacon requests on each channel (Algorithm~\ref{alg:probech}). Receiving a beacon frame in response to a beacon request indicates that there is a network on the current channel (note that generally only coordinators and routers reply to beacon requests, end-devices do not). Subsequent passive scanning rounds can then be limited to these active channels (line~\ref{alg:active:passive} in Algorithm~\ref{alg:active}), in order to detect any further devices that did not respond to active scanning (i.e., end-devices).

%We differentiate between passive and active Zigbee scan. The probe frame type which is sent as part of the active Zigbee scan (Algorithm~\ref{alg:probech}) is a \emph{beacon request}. Beacons are returned in response to beacon requests.
%, which are sent as part of the active scan. 
%That being said, Zigbee coordinators and routers reply to beacon requests but end-devices do not. Several devices in our Zigbee network experiments play the role of a router. Therefore, an active Zigbee scan discovers many devices in the first round of a channel scan. 

\subsection{Bluetooth Low Energy (BLE)}

Bluetooth LE~\cite{BluetoothCoreSpec5} is a popular short-range wireless protocol on the 2.4GHz ISM band. Its physical layer comprises 40 channels (0-39), three of which are so-called \emph{advertising channels} which are used to broadcast device information using \emph{advertising packets}. Bluetooth LE operates on 40 RF (radio frequency) channels in the 2.4GHz band. Each channel is 1\,MHz wide with center frequency $f_c = 2402 + 2k$\,MHz where $k=0...39$. The 40 RF channels are mapped to either data channels or advertising channels (see ~\cite{gupta2016inside}).
The advertising channels are centered at 2402\,MHz (channel~37), 2426\,MHz~(38), and 2480\,MHz~(39) to ensure coexistence with other wireless protocols, i.e. minimize interfering with the most populated Wi-Fi channels 1, 6, and~11.
%\subsection{BLE traffic analysis}
% paragraph below explains why we don't scan all 40 channels (why we haven't built such receiver)

\textbf{Address Information.} Advertising BLE packets contain two address-related data fields in the packet structure of their most common packet types: the \emph{access address} and the \emph{advertising address} (Fig.~\ref{fig:ble-frame-structure}. We use the advertising address (AdvA, 6 bytes long) to enumerate BLE devices. For advertising, the access address is set to the constant \texttt{0x8E89BED6} and is used as a sync word for frame synchronization. As it is the same for all advertisers, it cannot be used for device enumeration. %The advertising address (AdvA) is 6 bytes long and can be used to enumerate BLE devices, see~Fig.~\ref{fig:ble-frame-structure}.

%\subsection{Limitations}
The advertising addresses of the BLE devices tested in this work did not change over time. Hence, we focused on identifying scanned devices by their advertising address. Note, however, that some devices, such as Apple devices, randomize their advertising addresses over time~\cite{gupta2016inside,becker2019tracking}.
%In fact, random addresses are privacy feature in BLE which may be implemented by vendors \cite{gupta2016inside}.% Some BLE devices will have a new advertising address generated randomly (anti-tracking address randomization) and some will change it in more predictable manner such as incrementing by one.
 In such cases, to infer the identity of BLE devices, one could use the data payload of BLE advertising messages, which include device identifiers, counters, or battery levels~\cite{celosia2020discontinued}.
%Note that our scanning times of BLE devices are much faster than the typical randomization times of such devices (typically on the order of~15 minutes).
%about an order of However, not all packets sent from a device will have this identifying string in clear text - which would make this kind of BLE device enumeration slower.

\textbf{Implementation.} Data channels are used for communication after a connection has been established, whereas advertising channels are used between devices that are in range to discover one another and exchange metadata. Therefore, {\tt IoT-Scan} only scans the three Bluetooth LE advertising channels (i.e., there is no need to monitor data channels).
%can be scanned with Algorithm~\ref{alg:passive} to discover Bluetooth LE devices without the need to monitor all data channels.

%Furthermore, receiving BLE packets on data channels "blindly" as a sniffer who isn't part of the connection and doesn't know the access address (created randomly for each connection) and channel hopping pattern is difficult, whereas receiving packets on advertising channels is easier, as the advertising access address is constant (\texttt{0x8E89BED6}).

Advertising packets are sent on all three advertising channels for any given advertising event. This redundancy makes device discovery more resilient in cases where some of the channels experience interference. This means that scanning for BLE devices on any one of the three advertising channels is as good as a multi-channel scan (sequentially scanning each advertising channel), a fact that we also verified experimentally.
%We ran an experiment to show that device discovery times do not change significantly by scanning on either on of the BLE advertising channels. % Comparison of the single channel BLE scans for channel 37, 38, and 39 can be seen in Figure \ref{fig:ble_val}.

\subsection{LoRa}
\label{sec:lora}
\begin{figure}[t]
    \includegraphics[width=\linewidth]{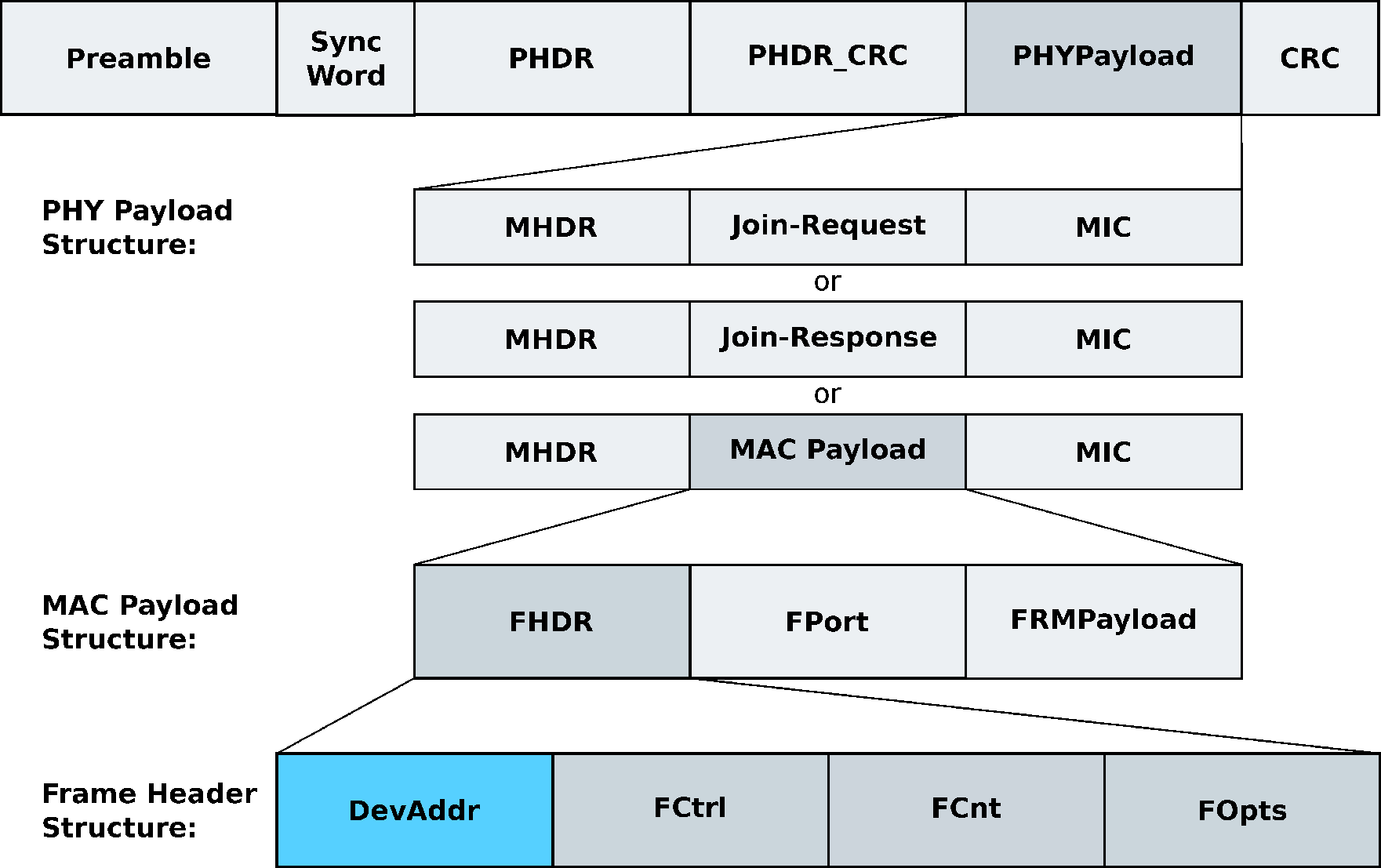}
    \caption{The LoRaWAN packet structure~\cite{lorawan-spec}.}
    \label{fig:lorawan-packet-structure}
\end{figure}

LoRa is a proprietary physical layer wireless protocol powering network layer protocols, such as  LoRaWAN~\cite{lorawan-spec} long-range wide-area networks (LP-WAN) and Sidewalk~\cite{amazon-sidewalk}.
%In the USA, LoRaWAN operates on uplink (towards gateway) and downlink (towards end-device) channels.
LoRa defines all supported modulations and physical layer signaling. On the other hand, LoRaWAN defines a subset of all possible modulations and signal parameters, such as frequency allocations and channel widths. The physical layer of the LoRa protocol was patented by Semtech\footnote{\url{https://www.semtech.com/}}, and the specification of LoRaWAN is governed by the LoRa Alliance~\cite{lorawan-spec}.  Work by security researchers have yielded SDR implementations of the physical LoRa layer~\cite{robyns2018multi,Tapparel2020}.

LoRaWAN uplink channels consists of 64 125\,KHz wide channels (centered around $f_c = 903.2 + 0.2k$\,MHz where $k=0...63$) and 8 500\,KHz wide channels (centered around $f_c = 903 + 1.6(k-64)$\,MHz where $k=64...71$). LoRaWAN downlink channels consists of 8 500\,KHz wide channels (centered around $f_c = 923.3 + 0.6k$\,MHz where $k=0...7$)~\cite{LoRaWANRegional}.
%The 125\,KHz uplink channels have center frequency $f_c = 903.2 + 0.2k$\,MHz where $k=0...63$. The 500\,KHz uplink channels have center frequency $f_c = 903 + 1.6(k-64)$\,MHz where $k=64...71$. The 500\,KHz downlink channels have center frequency $f_c = 923.3 + 0.6k$\,MHz where $k=0...7$ \cite{LoRaWANRegional}.
%Note that LoRa traffic direction is asymmetrical where most data flows from end-devices upstream to the network.

%that cover most of the protocol's physical layer to make common receiver/transmitter LoRa configurations possible on a GNU Radio based SDR platform, such as a spreading factor of 7, bandwidths of 125 or 500\,KHz, explicit header mode, etc.) 

%\color{red}YoLink devices we tested on 900MHz band adhere to LoRaWAN specification but it may not work with other off-the-shelf LoRaWAN devices.\color{black}
%LoRaWAN in North America defines 64 uplink channels (from 902.3 to 914.9MHz) that are 125\,KHz wide, 8  uplink channels (from 903MHz to 914MHz) that are 500 \,KHz wide; and another 8 downlink channels (from 923.3MHz to 927.5MHz) that are 500\,KHz wide, totaling 80 channels.

\textbf{Address Information.} 
%The MAC payload header of LoRa packets contain a network-specific sync word right after the preamble, and a device address, see Fig.~\ref{fig:lorawan-packet-structure}.
%, \color{red}and a Network identifier (NwkID) and sync word after the preamble\color{black}~ (Figure~\ref{fig:lorawan-packet-structure}).
We use the third byte after the sync word to enumerate the LoRa devices under test.
%which all happen to be made by YoLink \cite{yolink}.
Indeed, from the traffic we collected, the value of the third byte consistently changed between four values, corresponding to the IDs of the four YoLink devices under test. Incidentally, this third byte of the payload is part of the 32-bit device address (DevAddr) as specified in LoRaWAN frame format~\cite{lorawan-spec}. Furthermore, the first byte of DevAddr is used as a network identifier (NwkID), which is fixed for all devices in the same network. In this context, a network consists of a LoRa gateway and end-devices connected to that gateway. %To enumerate YoLink devices we only use the second byte of the DevAddr for convenience. %Because only the second byte changes from device to device.

% In addition to DevAddr, the sync word could be used to distinguish between different private LoRa networks. The sync word is 2 symbols long, appears directly after the variable-length preamble, and is primarily used for frame synchronization \cite{Tapparel2020}, as the name suggests. Private networks can have custom/private sync words which effectively filter out packets from outside of the network. LoRa differentiates between private (\texttt{0x1424}) and public (\texttt{0x3444}) sync words~\cite{sx1261}. However, LoRa developers are free to choose other custom values for a sync word.

%Frames containing the public sync word should be compliant with the LoRaWAN specification, whereas private LoRa networks can use custom packet structures. Sync word values containing \texttt{0x00} are forbidden in deployed networks and can only be used in testing.

\textbf{Implementation.} 
%While LoRa typically operates on sub-Gigahertz unlicensed bands (900MHz band in USA), Semtech recently released a LoRa transceiver for the 2.4GHz band~\cite{Semtech2.4Ghz}, which will enable LoRa-based communication products to coexist on the same band with the aforementioned wireless protocols. As the LoRa 2.4GHz development module is not restricted to particular frequency channels and the 2.4GHz band does not have an assigned channel map as sub-Gigahertz band does, we arbitrarily use the following channel map for 2.4GHz LoRa: $F_c = 2400 + k$\,MHz for channel numbered $k=0...80$.
In our implementation, we scan LoRa devices listed in Table~\ref{tab:testeddevices} using Algorithm~\ref{alg:multiproto}.
%All our devices belong to Yolink Smart Home \cite{yolink}.
YoLink's website does not specify whether or not their devices are LoRaWAN compatible. %A reference~\cite{yolinklorawanrumor} claims they are, 
However we found that the sync words used by YoLink devices are different from the public sync word (\texttt{0x3444}) - needed for LoRaWAN compliance~\cite{lorawan-spec}. %Regardless, {\tt IoT-Scan} can enumerate LoRa and LoRaWAN-compatible devices, because our implementation allows us to configure the LoRa's sync word, the bandwidth, the center frequency, the bitrate, and other parameters.

We observed that our YoLink devices only used two channels: 910.29\,MHz for the end-devices (uplink) and 923.29\,MHz for the gateway (downlink). Note that the uplink frequency is located between two official LoRaWAN frequencies. The YoLink's downlink frequency (923.29\,MHz) is very close to the LoRaWAN's downlink channel zero (923.3\,MHz). YoLink's downlink bandwidth is 125\,KHz and not 500\,KHz, as defined by LoRaWAN. YoLink's uplink bandwidth of 125\,KHz agrees with LoRaWAN specification. Given that YoLink does not exactly follow LoRaWAN's frequency allocations and channel widths, this gives further proof that YoLink devices are not LoRaWAN compatible.

%Instantaneous bandwidth used in multiprotocol scan was 8\,MHz.

The major challenge in receiving any Yolink traffic initially was in determining the network's custom sync word, because the default SDR LoRa receiver~\cite{Tapparel2020} accepts only sync words with a value of zero. Sync word values containing \texttt{0x00} are forbidden in deployed networks and can only be used for testing. We overcame this challenge by modifying the LoRa receiver of~\cite{Tapparel2020}. Our implementation allows one to promiscuously listen for all sync words, as well as configure the bandwidth, the center frequency, the bitrate, and other parameters.
A key advantage of scanning LoRa using an SDR implementation is that all sync words can be monitored simultaneously, whereas certified LoRa transceiver chips are programmed to receive a specific sync word.
%All tested YoLink 
 %Enumerating LoRa devices on 2.4GHz was done in a similar fashion using LoRa 2.4GHz module emulating LoRa traffic.
%After finding the network's sync word we again modified the LoRa receiver to selectively look for this sync word which improved the error rate compared to promiscuous receiver. Finally, we use one byte corresponding to the NwkID to enumerate Yolink LoRa devices. Incidentally the values of the four NwkID-s, corresponding to the four Yolink LoRa devices, were close to each other: 11,12,13, and 14.

 %Inspecting number of packet headers from individual Yolink device has revealed a single byte bitfield which stays fixed, indicating its local network address. 
 %Knowing the device address location (address byte offset) in a packet, we can enumerate Yolink LoRa smart devices.

% Emulated LoRa traffic on 2.4GHz band is enumerated using single byte at the end of the frame which we designate to represent the node id.

\subsection{Z-Wave}

\begin{figure}[t]
    \centering
    \includegraphics[width=\linewidth]{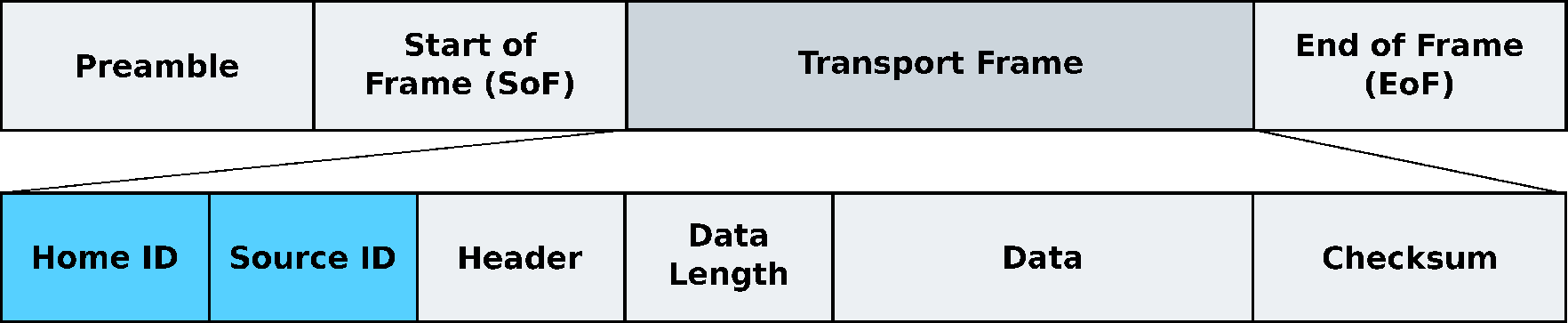}
    \caption{Z-Wave physical (PHY) and medium access control (MAC) layer frame structure \cite{9959}.}\
    \label{fig:zwave-frame-structure}
\end{figure}

Z-Wave is another proprietary physical layer wireless protocol, based on the ITU-T G.9959 specification~\cite{9959}. It is used in smart home applications, most notably Ring Home Security Systems.%~\cite{ring}. 
%Similar to Zigbee it uses AES 128 encryption in their new Security 2 (S2) framework. Z-Wave network supports 232 devices while Zigbee theoretically supports 65k devices. 

In the US, Z-Wave operates on the 900\,MHz ISM band and comes in a few physical layer (PHY) variants, most importantly differentiated by its center frequency and bit rates (channel widths): 908.4\,MHz at 9.6\,Kbps (R1) and 40\,Kbps (R2); 916\,MHz at 100\,Kbps (R3). Z-Wave long range PHYs at 912MHz and 921\,MHz~\cite{zwave-longrange} are less common. For Z-Wave devices listed in Table~\ref{tab:testeddevices}, {\tt IoT-Scan} discovered traffic on the R2 and R3 PHYs, with corresponding channel widths of 40\,KHz and 100\,KHz respectively.
%%From the specification (Fig.~\ref{fig:zwave-frame-structure}), the first three rates are denoted as R1, R2, and R3.

\textbf{Address Information.} Z-Wave packets contain two address-related data fields: the home (network) identifier, and the node (device) identifier. We use the single byte Source ID~\cite{badenhop2017z} to enumerate Z-Wave devices.
Z-Wave supports up to 232 devices per network, hence this byte is sufficient to distinguish between devices on the same network. The Z-Wave primary controller (Z-Wave gateway) has a Source ID of 1. A Z-Wave device which has not been connected to a controller must use a Source ID of 0 before obtaining an actual non-zero Source ID.
The Z-Wave network identifier or Home ID consists of 4 bytes that precede the node ID, as shown in Fig.~\ref{fig:zwave-frame-structure}. 
%The Home ID is the same for all devices (pre-programmed in the Z-Wave primary controller)
%in our network of Ring devices and it does not change over time.
In the case of multiple overlapping networks, the Home ID can be used to distinguish between devices on different networks.  
%However, each Z-Wave device has its own unique node ID. 

%After obtaining such node ID, it stays fixed. 

% Z-Wave does not have an equivalent of beacon request as in IEEE~802.15.4 for discovering nearby networks and devices~\cite{hall2016z}. In order to communicate with the Z-Wave network one needs to know the home ID. Z-Wave devices will only reply to packets containing their own home ID. Thus, before running any potential Z-Wave active scan, a passive scan identifying the home ID is required first.

% The Z-Wave transmissions of all our Ring devices occurred in unison every hour or so. Hence, the passive Z-Wave scanner needs an hour to find the home ID of the network. But shortly after the home ID is discovered, all Z-Wave devices have been discovered anyway, assuming they are transmit roughly at the same time. Therefore, Z-Wave active scan is not  very useful in enumerating our particular network of Z-Wave devices.

\textbf{Implementation.}
R1 and R2 Z-Wave PHY implementations are available from Picod et al.~\cite{sdr-pentest} and scapy-radio~\cite{scapy-radio}. We built the R3 Z-Wave PHY flowgraph based on the existing R2 PHY implementation, the only difference is that the bitrate/sampling rate has to be 2.5 times larger.
{\tt IoT-Scan} discovers Z-Wave devices by listening on the 908.4\,MHz and 916\,MHz channels.
%with Z-Wave R2 and R3 PHYs respectively.
%The PHY layer of the Z-Wave packets is not encrypted, specifically home ID, source, and destination IDs are in clear text.
%In particular, we used the source IDs to enumerate devices. 
We then enumerate the Z-Wave packet's source IDs with a custom made Z-Wave Wireshark dissector~\cite{zwave-dissect}. %Note that we do not analyze Z-Wave and LoRa transmission statistics due to its roughly periodic transmission pattern (devices transmit once every hour or so).

% moved Zigbee device table to earlier section for better layout 
\section{Experimental Evaluation}
\label{sec:exp}
In this section, we perform an experimental evaluation of the scanning algorithms of {\tt IoT-Scan}. We detail SDR implementation aspects, the experimental set-up (including the list of tested devices), and the experimental results.

% \subsection{computational load}
% receiver constraints: CFO, SCO

\subsection{Algorithm Implementation}
\label{sec:algo-imp}

\label{sec:gnuradio-processing}

The algorithms described in Section~\ref{sec:algorithms} do not go into details of SDR-level implementation. This section provides additional detail on the implementation of these algorithms. 
The main software components consist of GNU Radio 3.8~\cite{gr} and Scapy-radio 2.4.5~\cite{scapy}.
%and latest Scapy-radio's ~\cite{scapy-radio} GNURadio flowgraphs.
% Major software engineering effort was put toward physical layer radio protocol implementation changes to existing GNU Radio flowgraphs based on Scapy-radio~\cite{scapy-radio}.
Scapy-radio is a pentest tool with RF capabilities, a modification of the Scapy toolkit~\cite{scapy}. %, that supports packet manipulation, scanning, tracerouting, probing, network discovery, and more.
Note that Scapy-radio currently supports GNU Radio version 3.7, and therefore we had to port several
blocks to version 3.8.
%(2) cloning existing R2 Zwave PHY implementation to make R3 PHY.
%It is worth noting that our initial implementation used GNU Radio version 3.7 whereas our ported versions of the flowgraphs run in version 3.8. 
A significant feature improvement in GNU Radio 3.8 is the default automatic gain control (AGC) supported by the USRP Source block, which improves receiver reliability.

\subsubsection{Flowgraph control}
We implement the scanning algorithms described in Section~\ref{sec:algorithms} in Python. Signal processing parameters, such as the SDR center frequency, the channel frequency offsets, and the channel bandwidths, are managed by a GNU Radio flowgraph.
%The state of the flowgraph includes parameters that define a channel (as mentioned in Section~\ref{sec:algorithms}): 
The GNU Radio flowgraph is imported from the main application as a Python module and is controlled with its native Python API.
%can be called from the main python algorithm as a separate process or it can be imported as another python module. In the former approach, one may use xmlrpc or alternative tool to change the state of the flowgraph during its execution (dynamically).
This allows for dynamic control of flowgraph parameters during the runtime of the flowgraph.
%by accessing their respective setter methods directly.
Controlling the flowgraph in this way is crucial for correct time-keeping of the experiments, as it allows to compensate for flowgraph startup delays of several seconds resulting from the initialization of the USRP hardware driver library UHD \cite{uhd}, which sets up the SDR hardware before the flowgraph can be executed.

\subsubsection{Signal processing} \label{sec:channelization} \label{sec:signal-processing}
The process of converting an unfiltered full-bandwidth signal from an SDR source into the receive chain (i.e, the sequence of DSP blocks connected serially starting with radio source, demodulator, filter, and clock recovery) of a particular protocol is referred to as \emph{channelization} \cite{channelization}. Channelization comprises three signal processing steps: frequency translation (from the center frequency of the raw radio signal to the center frequency of the desired channel),
channel filtering (filtering out other protocols and potential interference),
and re-sampling (down conversion) to reduce the computational load.

Channelization is particularly important in multiprotocol scanning, since it reduces the computational complexity of the receivers by reducing the sampling rate. %Single protocol scans may run with higher-than-required sampling rates without dropping samples and overflowing. However, 
Multiprotocol scans require parallel decoding of two or more receive chains which can overwhelm the capabilities of a typical host computer if the processing chain in the flowgraph is not correctly optimized.

%\color{red}
Reducing the sample rate relies on Nyquist theorem, which dictates that the sample rate of a signal be at least twice the signal's bandwidth, in order to not lose any information (in other words, to not have signal aliasing or ambiguity in reconstructing the original signal that was transmitted). 
%In practise, the radio signal is over-sampled by some factor (e.g. 4 or 8) to compensate for signal imperfections, i.e. to allow for signal parameter estimation algorithms (e.g. timing recovery, carrier frequency offset compensation etc.) to recover the signal properly.
Channel filtering reduces the total captured signal bandwidth down to the bandwidth of a single protocol channel  in a given received chain. The reduced sample rate matches the reduced bandwidth of the filtered signal, such that the Nyquist rate condition still holds after  channelization.

\subsection{Experimental Setup}

\begin{table*}[!htb]
    \caption{Tested IoT devices.}
    %The number next to each device name is shown in Figure~\ref{fig:zigbee-expsetup}.}
    \label{tab:testeddevices}.
    \includegraphics[width=.8\linewidth,trim=1cm 11cm 6.7cm 2.3cm, clip]{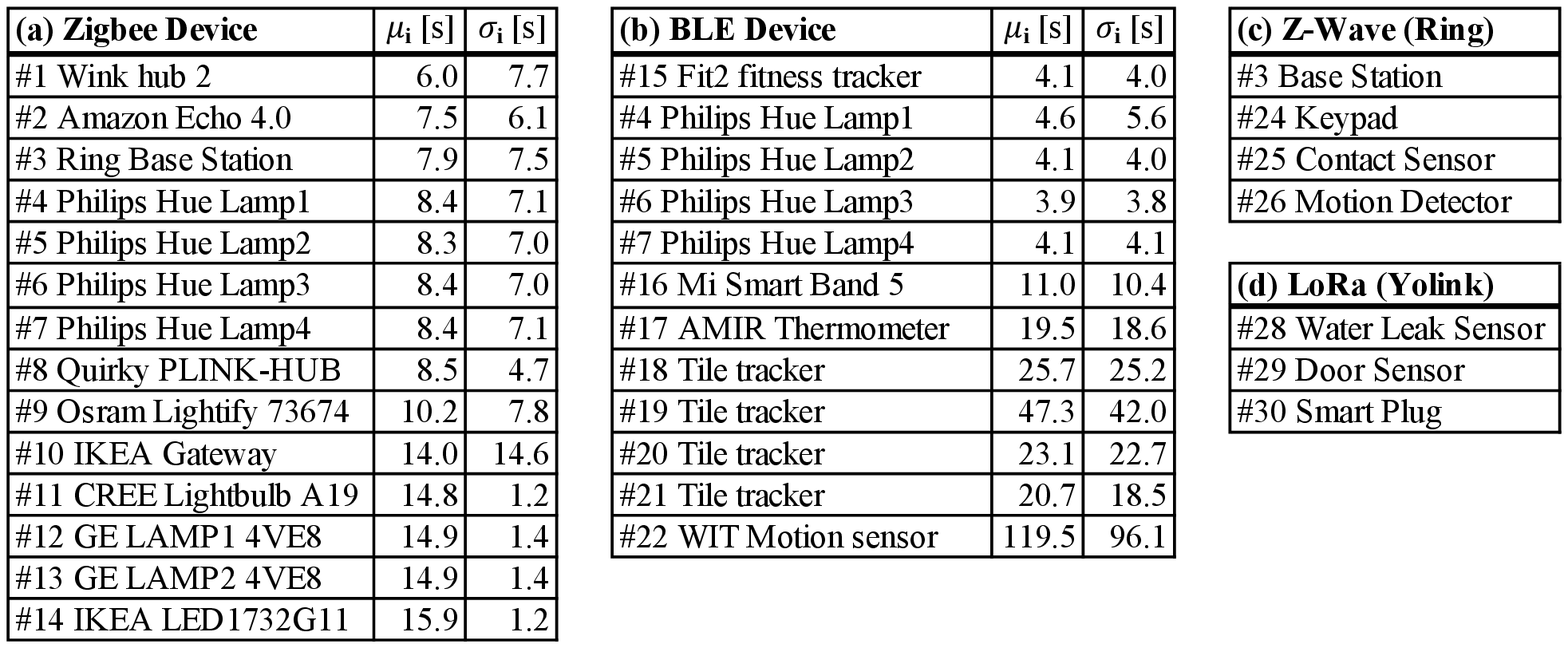}
\end{table*}

 We implemented all the scanning algorithms described in Section~\ref{sec:algorithms} on a single SDR device, namely a USRP B200 device~\cite{ettus-usrp-b200}, with a PC capable enough to handle data processing in real-time without dropping samples (i.e., overflowing buffers). Thus, all our experiments were run on a ThinkCentre 8 Core Intel i7 running Ubuntu 20.04.

%\paragraph{Tested Devices}

The devices used in the experiments are listed in Table~\ref{tab:testeddevices}. 
%These devices are used for the experiments described in the following sections. 
Traffic of the BLE and Zigbee devices were statistically analyzed to derive the parameters of the theoretical traffic model introduced in Section~\ref{sec:traffic-analysis}. Note that we did not analyze transmission statistics of low-power Z-Wave and LoRa devices due to their periodic transmission patterns (devices transmit once every hour or so).

\begin{figure}%[t]
     \centering
     \includegraphics[width=\linewidth,height=6.5cm,trim=0 1cm 0 0, clip]
     {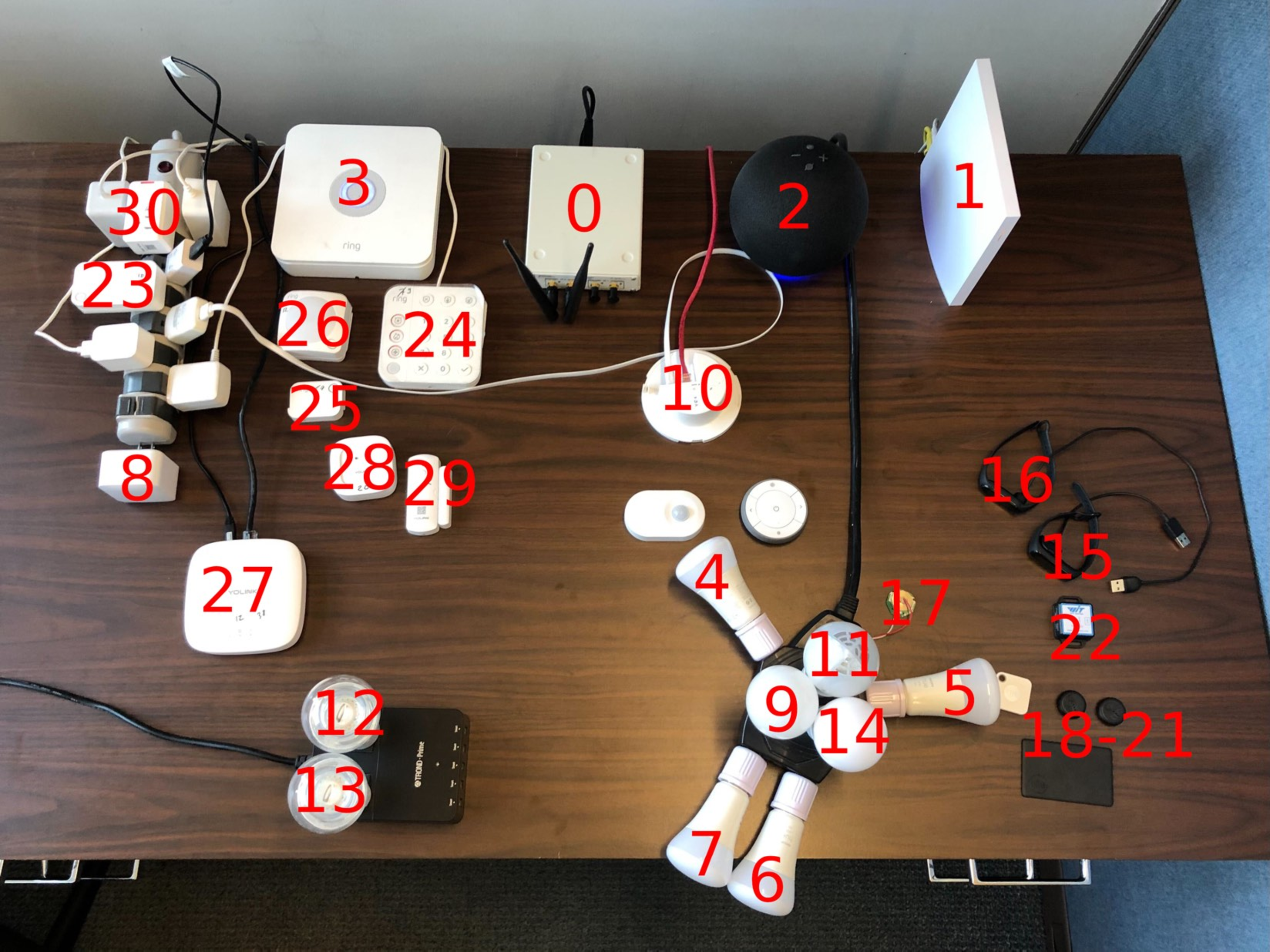}
 `   \caption{The experimental setup for the tested IoT devices, see Table~\ref{tab:testeddevices} for the numbering. The SDR is labeled with number 0. 
 %\color{red}TODO should have 3 groups for 3 active channels. Also show devices not association as single dots\color{black}
 }
     \label{fig:zigbee-expsetup}
 \end{figure}

% \begin{figure}[t]
%     \centering
%     \includegraphics[width=\linewidth,height=6.5cm,trim=0 1cm 0 0, clip]{figures/devices.png}
% `   \caption{The experimental setup for IEEE~802.15.4 target devices. 
% %\color{red}TODO should have 3 groups for 3 active channels. Also show devices not association as single dots\color{black}
% }
%     \label{fig:zigbee-expsetup}
% \end{figure}

All scanning experiments were based on IoT devices under our control. Any foreign device from the environment was filtered out. In order to only account for our devices, we initially  enumerated them with a passive scan inside an RF shielded box (Ramsey box STE3500) to determine their addresses. 

%To enumerate Zigbee devices we use 12 different short addresses (16bit), and 12 extended addresses (64bit). Note that a single Zigbee device has both a short and an extended address.
%, as well as PAN IDs when applicable (devices that respond to active scan).

%For 12 BLE devices, we identified 12 advertising addresses. No foreign LoRa and Z-Wave interference was observed in our laboratory, which made filtering network traffic on these protocols unnecessary.

%The scanning experiments are performed indoors with no other Zigbee traffic in the environment, to the best of our knowledge. Nevertheless, in order to have controlled and structured experiments we filer out any unknown packets coming from devices not under our control. In order to do that we manually enumerate associations of 16bit and 64bit addresses and device names (Table~\ref{tab:testeddevices}). Some devices may be associated with multiple address field types. Device names are human readable strings we used to label/enumerate each of tested hardware smart devices with compatible Zigbee radio. This way we only measure scanning performance based on devices under our control, any unknown device we don't control is filtered out before our analysis. The addresses of our devices do not change in short term

We conducted all scanning experiments using the default network configuration of the respective devices and protocols, i.e., by setting them up via their respective gateways and/or apps (wherever applicable) in preparation of any experiments.
%Zigbee Ikea lightbulb was connected to the Ikea motion sensor, Ikea switch/dimmer, and Ikea Gateway through the pairing procedure described on their website \cite{ikea-getting-started}.
During all the experiments, the tested devices (see Table~\ref{tab:testeddevices}) were in an idle state, i.e., not actively used by an operator. Manually operating devices in a way that generates network communication, e.g., actuating Zigbee lights via the Amazon Alexa smartphone app, would impact scanning performance.  We expect the results presented in this section to be conservative estimates of the scanning time, since generating additional traffic from the devices should speed up the discovery of the devices.   %However, ruling out sporadic smartphone user interaction gives a more reproducible upper bound on the discovery times.
%However, we didn't interact with the devices during scanning experiments to get more objective scan times that don't depend on the frequency of user interaction.
\begin{comment}
Additional network configuration consisted of the following:
\begin{itemize}
    \item The Amazon Echo 4.0 was paired to all smart light bulbs.
    \item Most BLE devices in our experiments come with an associated smartphone application, which was installed and the BLE device was paired to the same smartphone. Specifically, Philips Hue lights were connected to the Amazon Echo 4.0 using the Amazon Alexa app. Tile trackers, Fit2, and the Mi smart band were activated using their respective companion apps.% were activated using the Tile app; Fit2 fitness trackers were connected with the Galaxy Fit app; and the Mi smart band via the Mi Fit app.
    \item The YoLink LoRa devices were left in their default configuration after adding the end devices (moisture sensor, smart plug, door sensor) to the YoLink LoRa hub, using the YoLink app.
    \item The Z-wave devices (motion sensor, door sensor, keypad, and range extender) were paired to the Ring Security Base Station using the Ring app.
\end{itemize}
\end{comment}
All devices were located near the SDR, as shown in Fig.~\ref{fig:zigbee-expsetup}.

 Regarding the parameters of the algorithms, the channel dwell time (i.e, the scanning time of each channel in each round) was set to 1~second.
Exceptions are the channel dwell time 
during active scan of Zigbee (set to 0.2~s), and in Section~\ref{sec:chdwt} where we evaluate the impact of different channel dwell times on the performance of the algorithms. In each experiment, the total scan time was set to be long enough for all devices to be discovered.

When scanning each individual protocol, we set the instantaneous bandwidth parameter in accordance with the respective protocol's bitrate. Specifically, BLE's channel bandwidth was set to 1\,MHz, Zigbee to 2\,MHz, LoRa to 125\,KHz, and Z-Wave to 40/100\,KHz. When implementing multiprotocol scanning algorithms, we used wider bandwidth to fit the bandwidth of each protocol and channel spacing in between. Both the Zigbee/BLE and Z-Wave/LoRa and  multi-protocols experiment used 8\,MHz of bandwidth, which in each case was sufficient to simultaneously capture channels from each protocol, as discussed in the sequel.

%For example, note that 8\,MHz is large enough for parallel reception of some BLE and Zigbee channels. 
%channel 37 and Zigbee channel 11. 
%The Z-Wave/LoRa experiment also used 8\,MHz bandwidth, which is sufficient to capture two popular Z-Wave channels at  908.4\,MHz and 916\,MHz and a LoRa channel at 910.23\,MHz. 

\subsection{Results}
\label{sec:results}
In this section, we discuss experimental results of the scanning algorithms. The figures show the sample means and 95\% confidence intervals of the order statistics of the discovery time of the $n$-th device (see Eqs.~(\ref{eq:sample_mean}) and~(\ref{eq:conf_int})). Each point represents an average over 10 experiments with identical parameters.

%The settings of the channel lists and instantaneous bandwidth are detailed in each of the following subsections.

%along with the comparison with respective models where applicable.
% \label{sec:zigbee}

% \color{red}
% \begin{enumerate}
%     \item Passive Zigbee / BLE + their theoretical models
% % Implementation based on scapy-radio, relate to other related work how re we different beter?
% %   Gnuradio, socket, xmlrpc changing variables dynamically to avoid 6s UHD start delay
% % Comment on details in python script, so that people could reproduce it w/o python script
% % How to instantiate gnuradio flowgraph in python: 1) call as process 2) call from within python
% % How do we scan: how many loops, loop for channel swithcumh
% % Loop for reading from socket
% % Describe saving packets to pcaps
% % Extend if packets are collected recently, to reduce chance of truncated packets

%     \item Active Scan Zigbee vs passive
%     \item Multiprotocol Zigbee/BLE vs two sequential scans
%     \item Lora vs Z-Wave
%     \item Dwell time
% \end{enumerate}

\color{black}
\subsubsection{Passive Zigbee and BLE Scans and Comparison with Theoretical Model}

We first evaluate the performance of the passive scanning algorithms (Algorithm~\ref{alg:passive}) for Zigbee and BLE devices, and compare those with the expected discovery times based on the theoretical model described in Section~\ref{sec:traffic-analysis}. 

% scripts: compare_multi.py zig_active/passive.py
\begin{figure}[t]
    \centering
    \includegraphics[width=.9\linewidth]{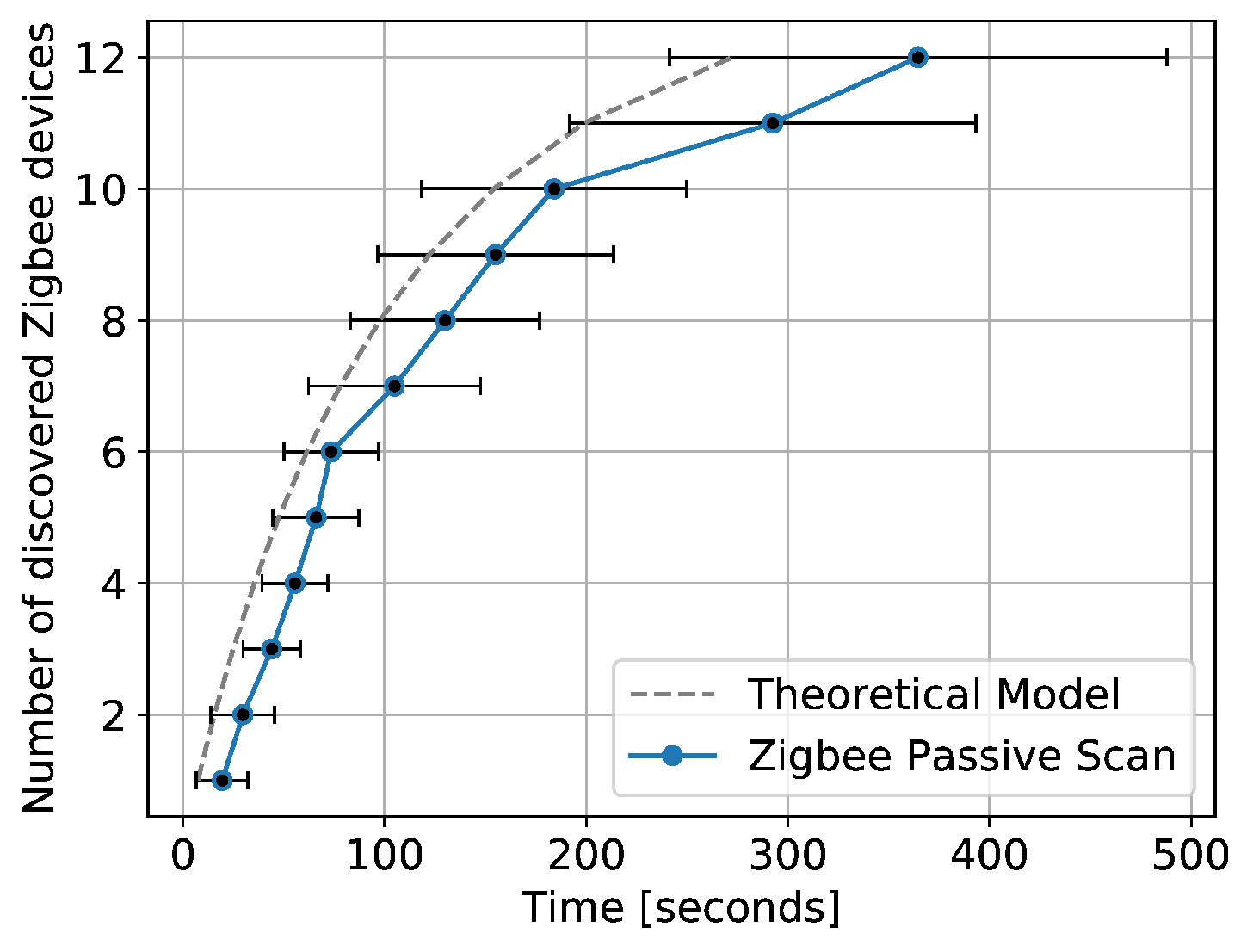}
    \caption{Zigbee theoretical model and experimental passive scan results. The 95\% confidence interval indicates a good fit.
    %\color{red}TODO: add data series for modeled expectations, see \texttt{traffic\_analysis\_zig.txt} in the repo\color{black}
    }
    \label{fig:zig-pas-mod}
\end{figure}

To build the theoretical traffic model (see Section~\ref{sec:traffic-analysis}), we measured device characteristics of our tested devices by running one long continuous scan of 100 minutes on every Zigbee channel and on every BLE advertising channel, in order to collect a baseline of traffic for each device. The traffic statistics are shown in Table~\ref{tab:testeddevices}. We set $\Delta t = 0.1$s, in Eq.~(\ref{eq:proba-single}) to compute $p_i$ for each device (this ensures that $\Pr(Z \geq 2)$ is negligible). We then use Eq.~(\ref{eqn:exp_tcn}) to compute the expectation of the order statistics of the discovery time of devices. Note that for Zigbee, we replace $p_i$ by $p_i/16$, since with Algorithm~\ref{alg:passive}, the SDR listens to only one out of the 16 Zigbee channels at a time.

Fig.~\ref{fig:zig-pas-mod} shows curves for the experimental results of Zigbee passive scanning and the theoretical model. 
%\color{red}
The model fits inside most of the 95\% confidence intervals. This shows that our passive scan implementation is close to the best performance possible, and our testbed has minimal packet losses. The deviation from the model could be attributed to interference (e.g., from Wi-Fi) and the fact that transmissions of some Zigbee devices are not memoryless.
%and the USRP's starting delay of a few seconds.
%\color{black}

%Of course, the performance will depend on the exact frequency of Zigbee packet transmissions per device.

\begin{figure}[t]
    \centering
    \includegraphics[width=.9\linewidth]{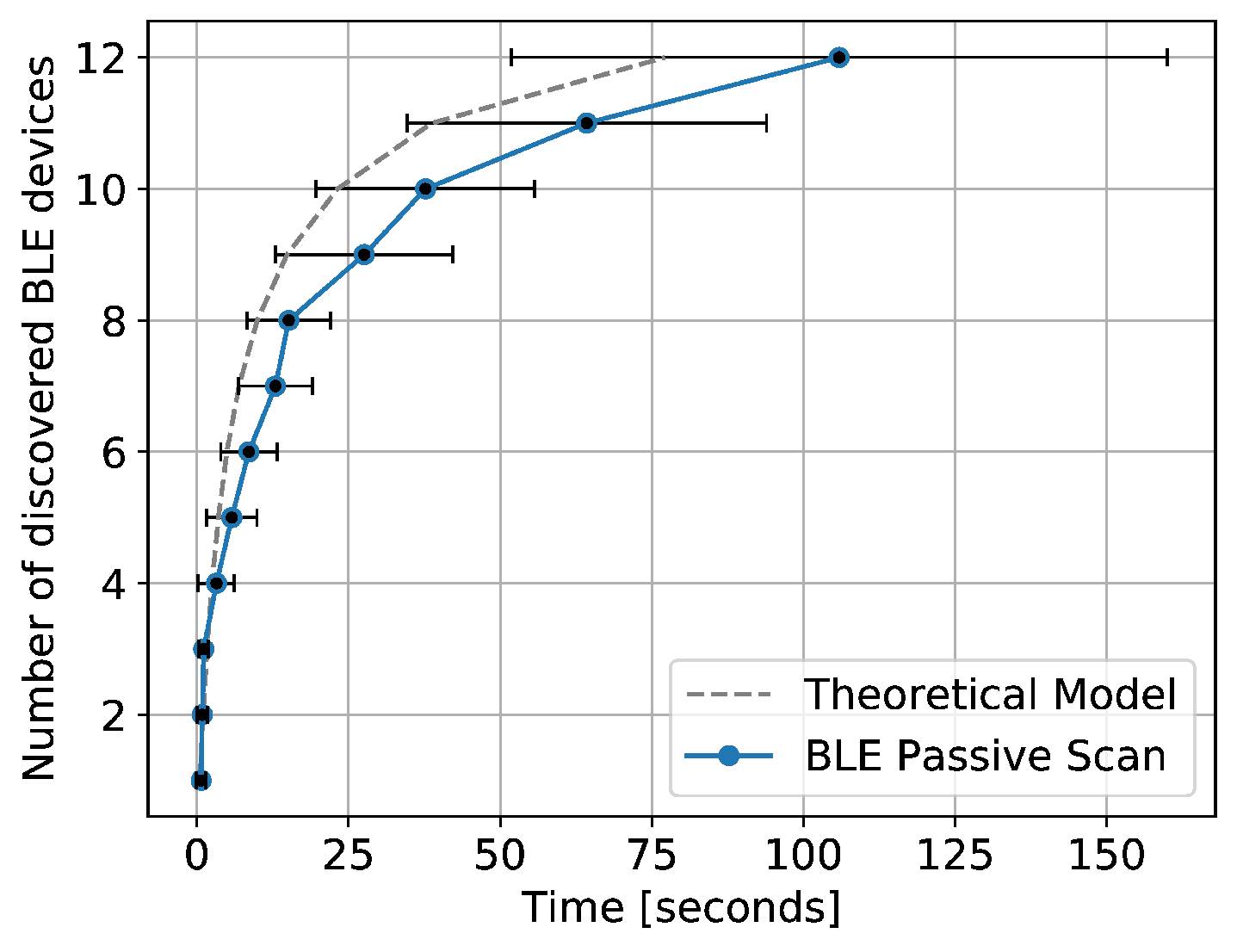}
    \caption{BLE passive scan results align closely with the theoretical model.}
    \label{fig:ble-pas}
\end{figure}

%\subsection{BLE}
Fig.~\ref{fig:ble-pas} shows experimental results for BLE passive scanning and the theoretical benchmark. The measured discovery times again fit the model well. 
Since all BLE advertising channels are equivalent, scanning is performed on channel 37 only. 
%This is supported by the device statistics in Table~\ref{tab:testeddevices}, which suggests that BLE advertising traffic can be assumed to be memoryless, as the model assumes.
Note that BLE device discovery can only be performed as a passive scan, since BLE does not allow for broadcast-type scan requests as performed in Zigbee. While BLE scan requests could be a useful active scanning technique for gathering additional device data, they are always directed scans, i.e., they require knowledge of the target device's address. 
%We did not analyze Z-Wave and LoRa transmission statistics due to the roughly periodic transmission pattern of the devices (devices transmit once every hour or so).

\subsubsection{Active Zigbee Scan}

\begin{figure}[t]
    \centering
    \includegraphics[width=.9\linewidth]{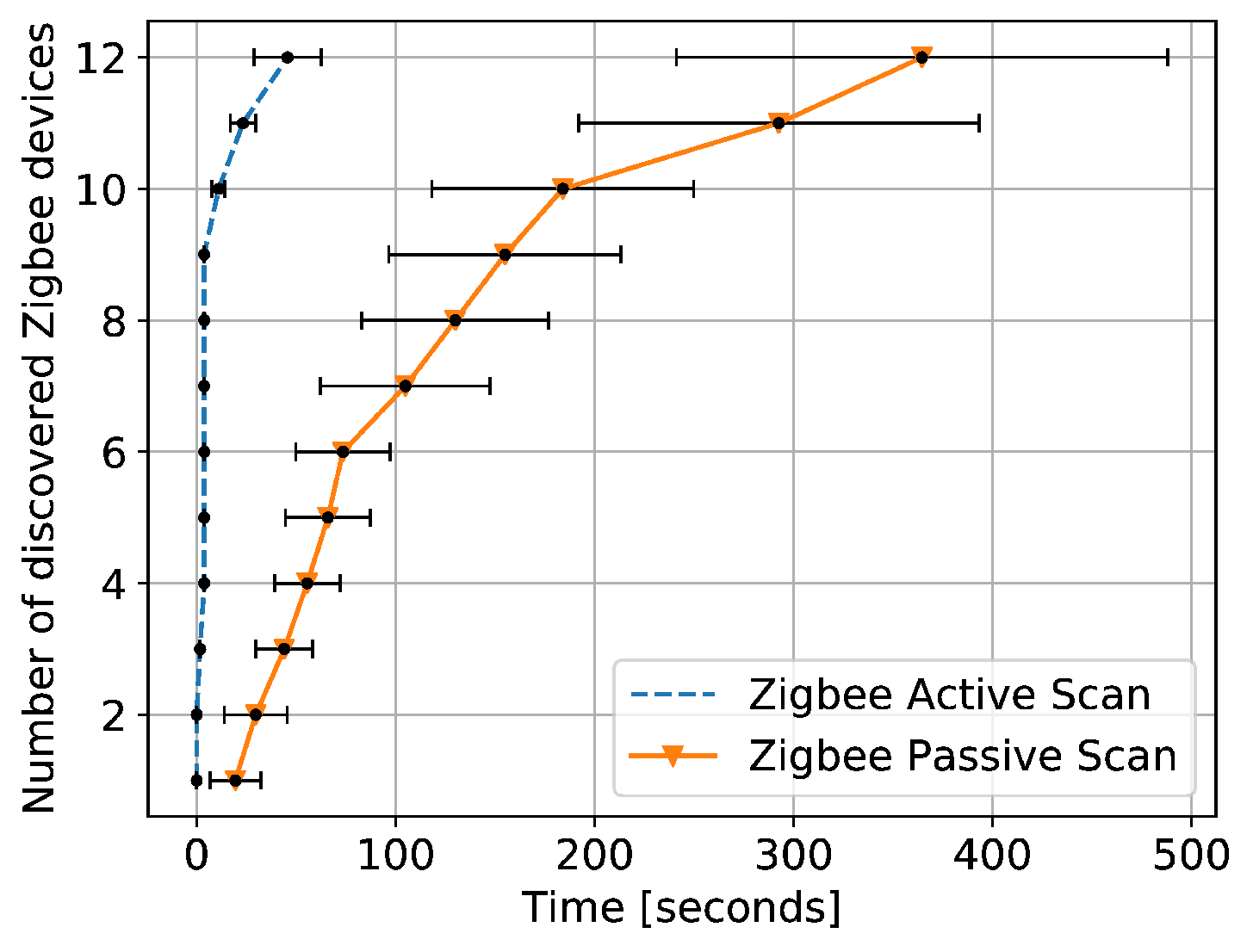}
    \caption{Zigbee active versus passive scan.
    %\color{red}TODO: add data series for modeled expectations, see \texttt{traffic\_analysis\_zig.txt} in the repo\color{black}
    }
    \label{fig:zig-act-pas}
\end{figure}

%Figure~\ref{fig:zig-act-pas} shows experimental results for Zigbee active versus passive scanning. 
We next evaluate the performance of active scanning (Algorithm~\ref{alg:active}) and compare it to passive scanning in the context of Zigbee. Fig.~\ref{fig:zig-act-pas} shows that the passive discovery of 12 Zigbee devices takes 365 seconds on average while  active Zigbee discovery takes only 46 seconds, i.e., a reduction of 87\% in the scan time. While active scanning discovers the 12 devices within one minute, passive scanning discovers only 4 devices within one minute.
%Within five minutes, the passive algorithm barely discovers 11 out of 12 devices in our test setup.

Note that Zigbee supports up to 64,000 nodes per network. It is conceivable that the improvement of active scan over passive scan would be even more significant with a larger number of nodes. %Most of the tested Zigbee devices are mains-powered routers, with the exception of a few battery-powered Zigbee end-devices such as the IKEA remote and motion sensor.
Zigbee routers and coordinators are typically continuously active and will reply to beacon requests, which contributes to the discovery of several devices almost immediately during the active scan, whereas end-devices are usually optimized for power saving, and may not respond to beacon requests. %, which makes active probing less effective on this type of device.
However, since end devices are on the same channel as their coordinator, limiting the second phase of the active scan to the known active channels significantly speeds up discovery by virtue of spending more time on each relevant channel. Among our tested devices, we have three Zigbee coordinators %: Amazon Echo 4.0, the IKEA Gateway, and a GE Link/Quirky hub. They
occupying three channels: GE Link/Quirky hub on channel 11, % (2405\,MHz),
IKEA Gateway on channel 15, % at 2425\,MHz (IKEA Gateway),
and Amazon echo 4.0 on channel 20. % at 2450\,MHz (Amazon Echo 4.0).
As a result, the second phase of the active scan (cf. Algorithm~\ref{alg:active}, line~\ref{alg:active:phase2scan}) cycles through 3 instead of 16 channels, shortening  the detection speed by a factor of roughly $16/3=5.333$ for the remaining end-devices.
%\color{red}The active scan algorithm (Algorithm~\ref{alg:active}) consistently detected three active channels, which were 11, 15, and 20. This is consistent with Zigbee's preference for low-interference channels, e.g., channel 15 and 20 do not have any overlap with the busy Wi-Fi channels 1, 6, or 11, and Zigbee channel 11 is on the outer edge of the 2.4Ghz band, with only slight overlap with Wi-Fi channel 1.\color{black}

\subsubsection{Zigbee and BLE Multiprotocol  Scan}

\begin{figure}[t]
    \centering
    \includegraphics[width=.9\linewidth]{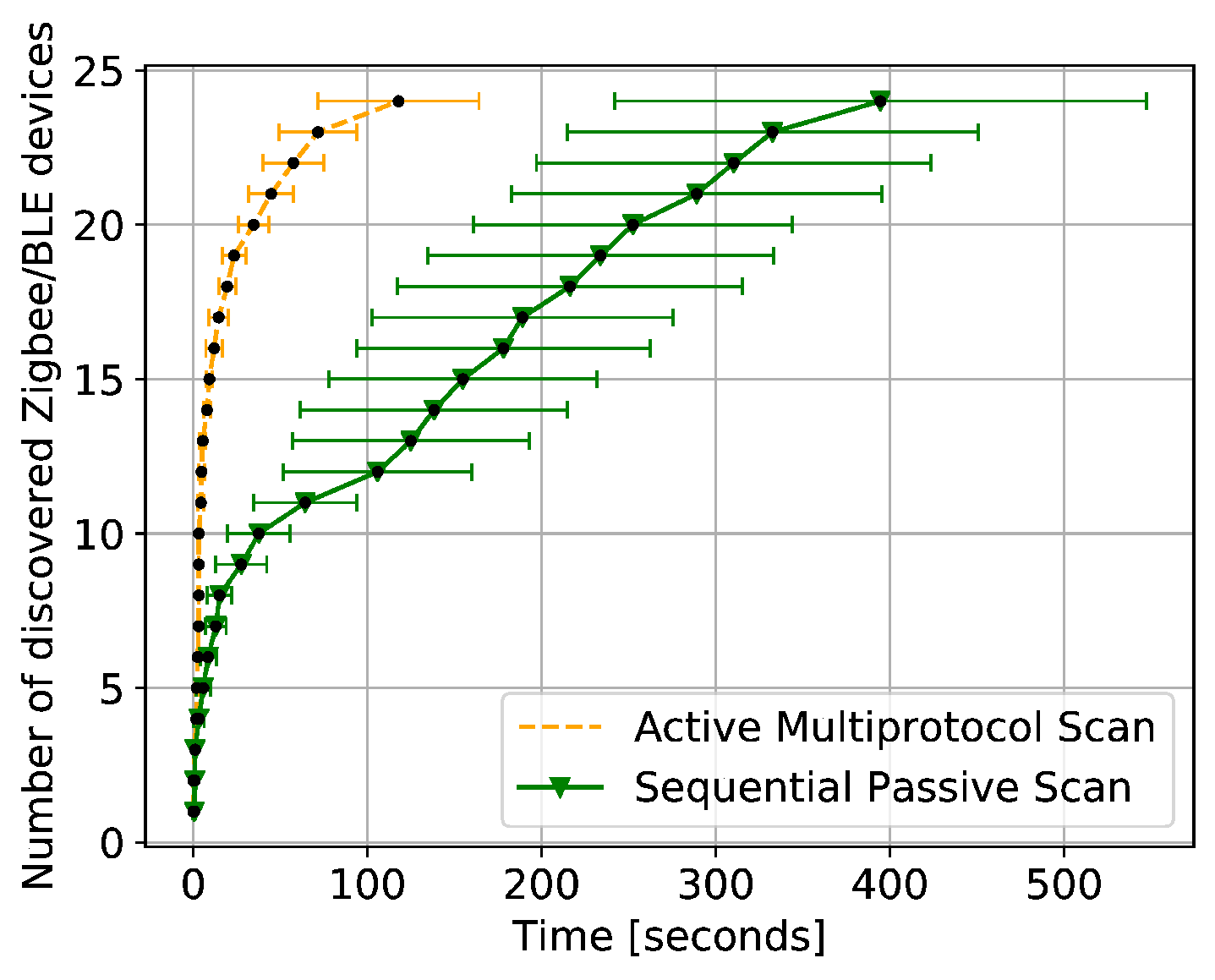}
    \caption{Zigbee/BLE multiprotocol active scan  
    vs.\ sequential passive scan.}
    \label{fig:multi-zigbee-ble}
\end{figure}

% \begin{figure}[t]
%     \centering
%     \includegraphics[width=\linewidth]{figures/zigble_model.pdf}
%     \caption{Active multiprotocol scan with zigbee and ble compared to the theoretical model}
%     \label{fig:multi-zigbee-model}
% \end{figure}
We next evaluate the performance of active multiprotocol Zigbee and BLE scan (Algorithm~\ref{alg:active-multiproto}) and compare it to sequential passive scan (Algorithm~\ref{alg:passive}). Sequential passive scan consists of passive BLE scan followed by passive Zigbee scan.
Sequential passive scan enumerates the 24 considered devices in 395 seconds on average, while active multiprotocol Zigbee and BLE scan takes 118 seconds on average, which corresponds to a 70\% improvement (Fig.~\ref{fig:multi-zigbee-ble}). Within 1 minute active multiprotocol scan discovers 22 devices while sequential scan discovers only 10.
%Within 5 minutes active multiprotocol scan discovers all the devices while sequential scan discovers 23 devices.
 Breaking down sequential passive scan into two: the first 106 seconds corresponds to a BLE passive scan, followed by 289 seconds of Zigbee scan, which is consistent with the results shown in Figs.~\ref{fig:zig-pas-mod} and~\ref{fig:ble-pas}.
%Stitching together Zigbee passive scan curve in Figure~\ref{fig:zig-pas-mod} and BLE passive scan in Figure~\ref{fig:ble-pas} piecewise matches the sequential passive scan, cross-validating our results.

The speed-up is achieved because of two aspects: active scan and multiprotocol scan. Zigbee active scan narrows the search down from 16 to only 3 channels. Multiprotocol scan supports reception of one Zigbee and one BLE channel in parallel. Note that parallel reception is possible only if the two channels fit within the instantaneous bandwidth. As mentioned earlier, the instantaneous bandwidth for multiprotocol scan was set to 8\,MHz. Three Zigbee active channels were identified, namely channel 11, 15, and 20. BLE has three well-known advertising channels, namely 37, 38, and 39. BLE channel 37 and Zigbee channel 11 can be received in parallel as well as BLE channel 38 and Zigbee channel 15. However, Zigbee channel 20 and BLE channel 39 are scanned separately since they do not fit within the same instantaneous bandwidth. 
% Figure~\ref{fig:multi-zigbee-model} shows comparison of experimental active multiprotocol scan and the corresponding theoretical model. Theoretical multiprotocol model assumes parallel Zigbee and BLE passive scan on known channels: 3 active Zigbee channels and 3 BLE advertising channels. The model does not take into account the Zigbee active scan which lasts for no more but a few seconds. The deviation from the model accumulates over time. Delayed discovery times of experimental results is indicative of small packet error rate due to insufficient computing power in running two protocol receivers.
% The instantaneous bandwidth of 8MSPS allows two of the three active Zigbee channels (11 and 15) to be scanned in parallel with the two corresponding BLE advertising channels (37 and 38).

\subsubsection{Z-Wave and LoRa Multiptotocol Scan}
%\color{red}
We next evaluate the performance of passive multiprotocol LoRa and Z-Wave scan on 900\,MHz band (Algorithm~\ref{alg:multiproto}) and compare it to sequential passive scan (Algorithm~\ref{alg:passive}). Passive multiprotocol scan consists of scanning each of 3 frequency channels (2 Z-Wave and 1 LoRa) in a round robin fashion. The passive scanning operation %takes place on the 900\,MHz band,
visits the LoRa and Z-Wave channel in a round-robin fashion, one at a time. Due to having 2 Z-Wave channels (908.4 and 916\,MHz) and only 1~LoRa channel (910.29\,MHz), Z-Wave has an  advantage in passive scanning.

\begin{figure}[t]
    \centering
    \includegraphics[width=.9\linewidth]{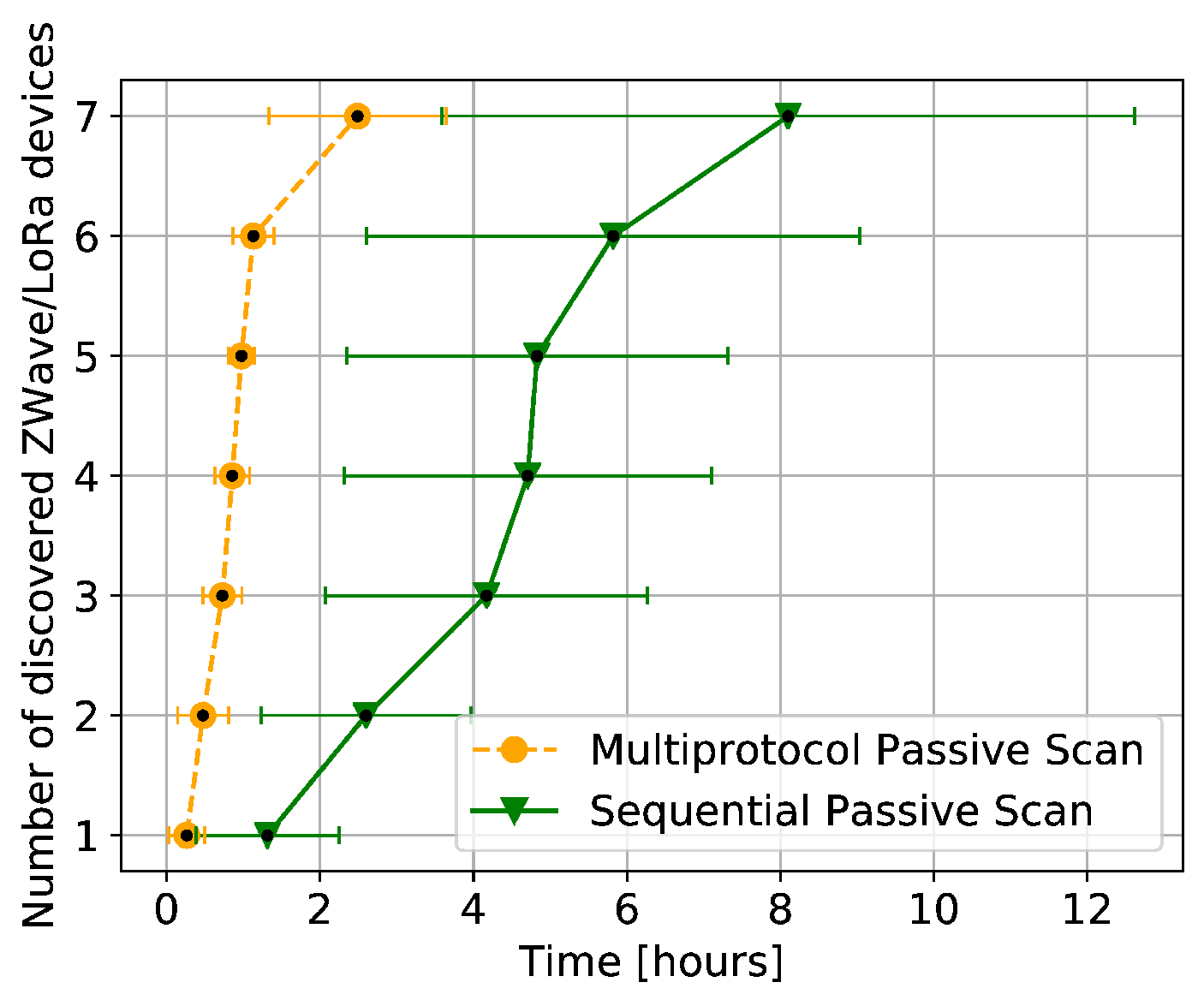}
    \caption{Multiprotocol passive scan (Z-Wave, LoRa)}
    \label{fig:multi-zwave-lora}
\end{figure}

Fig.~\ref{fig:multi-zwave-lora} shows that sequential LoRa and Z-Wave scan takes about 8.1 hours on average while multiprotocol Z-Wave and LoRa scan takes 2.5 hours, which represents a reduction of about 70\% in the discovery time.  Within a single hour passive scan discovers less than 1 device on average while multiprotocol scan discovers 5 out of the 7 devices. This significant speed-up is achieved because multiprotocol scan receives all three channels (from the two protocols) in parallel, namely
%However, within three hours multiprotocol scan discovers all 7 devices and single protocol scan finds only 2.
%The multiprotocol scan has three channel hops: 
908.4\,MHz (Z-Wave R2 PHY), 910.23\,MHz (LoRa uplink), and 916\,MHz (Z-Wave R3 PHY).%; 923.29\,MHz (LoRa downlink).
%Note that the total instantaneous bandwidth in this experiment was 2\,MHz for passive scan and 8\,MHz for multiprotocol scan.
%Scanning channels , as a second channel hop in multiprotocol scan, would require sampling rate too large to process all data in real-time on our PC. 
% Each scan trial was capped at 3 hours. We did not reliably scan more than 5 devices in that time which is why Fig.~\ref{fig:multi-zwave-lora} is capped at 5 out of 9 LoRa and Z-Wave devices listed in Table~\ref{tab:testeddevices}.

%While the sequential  scan hops over one LoRa channel, and on two Z-Wave channels, one channel at a time. Single protocol LoRa scan takes 4.2 hours (first 3 points on sequential passive scan in Fig.~\ref{fig:multi-zwave-lora}). Z-Wave single protocol then takes remaining 3.9 hours. Because the scan times for these two protocols, Z-Wave and LoRa, are roughly equal, the multiprotocol scan achieves good improvement.

%\color{black}

\subsubsection{Channel Dwell Time}
\label{sec:chdwt}

\begin{figure}[t]
    \centering
    \includegraphics[width=.9\linewidth]{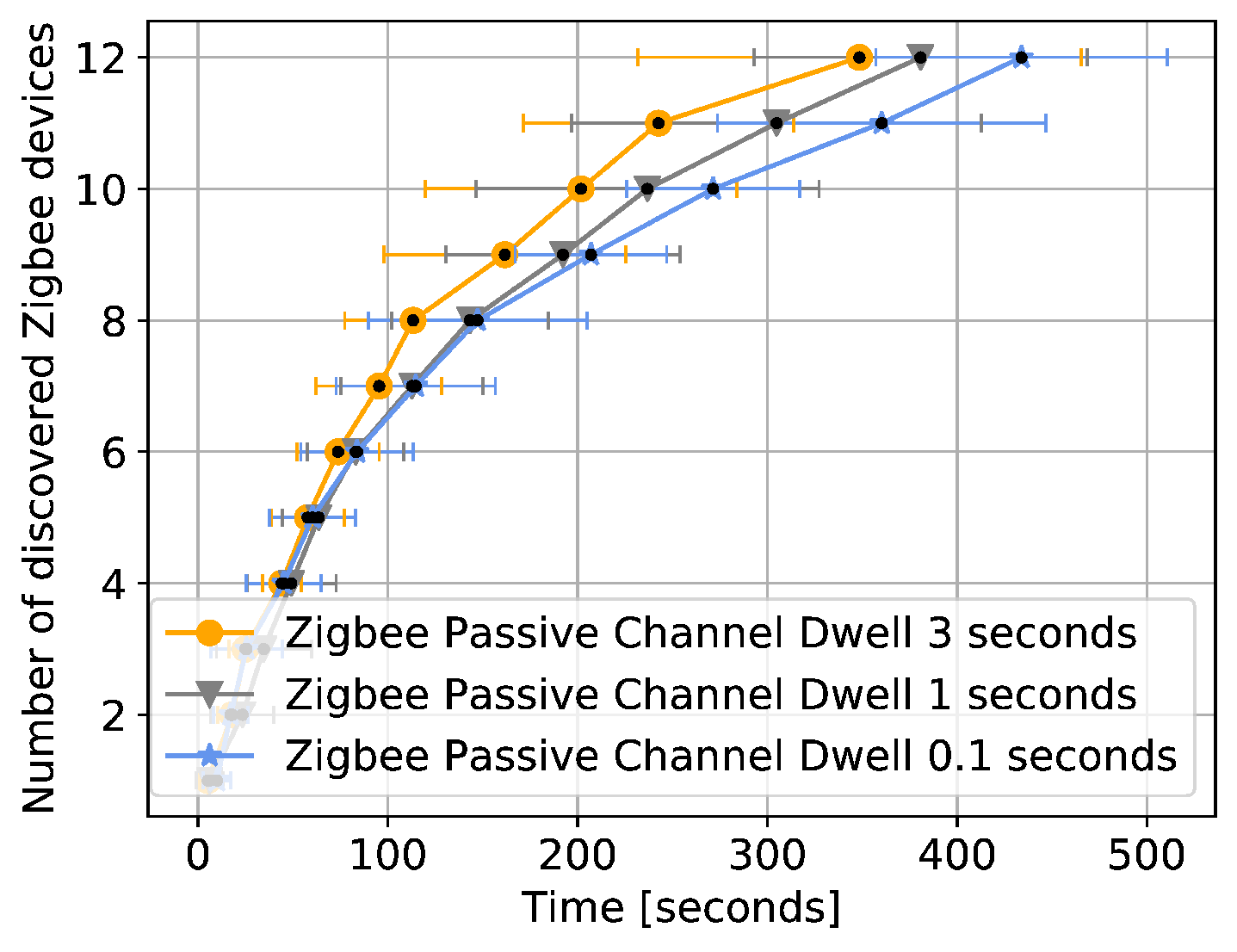}
    \caption{Zigbee passive scan discovery times for channel dwell times of 0.1, 1, and 3 seconds.}
    \label{fig:zigdwell}
\end{figure}

%The channel dwell time is an important parameter in all the algorithms that determine the listening duration on each channel hop.
We last examine the impact of properly setting the value of the channel dwell parameter, which is an input to several of the algorithms. Each trial measures the time to passively scan 12 BLE and 12 Zigbee devices. Passive Zigbee scan hops between 3 active Zigbee channels. A passive BLE scan involves channel hopping between three advertising BLE channels.
%Traffic scanned comes from our 12 BLE test devices.
Our experiments indicate that the scanning times do not differ significantly for channel dwell times of 0.1, 1, and 3 seconds. Fig.~\ref{fig:zigdwell} shows the channel dwell time experiment for Zigbee, but similar results hold for BLE.
%In our case, packet length averages z for Zigbee, b for BLE, and l seconds for LoRa, which is well under the channel dwell time of 1 second.
%(Figures \ref{fig:zig-val},\ref{fig:ble-val},\ref{fig:lora-val}).
Thus, in all our experiments, we chose a value of 1~second for the channel dwell times, the only exception being in the first phase of Zigbee active scan, where the dwell time was set to 0.2 seconds (which was sufficient to receive beacon responses and move on). 

\section{Conclusion}
\label{sec:conclusion}

We presented {\tt IoT-Scan}, a protocol-agnostic and extensible network reconnaissance tool for the Internet of Things that can be employed for network monitoring and security auditing. {\tt IoT-Scan} leverages the capabilities of SDRs to process multiple streams in parallel.  Accordingly, we introduced several scanning algorithms and evaluated them both theoretically and experimentally. Using the theoretical model, we showed that our implementation is efficient and achieves minimal packet loss in reception. 
%conditioned upon their frequency range, bandwidth, and processing capabilities of the host PC. 
We implemented multi-protocol, multi-channel scanning both on the 2.4GHz band for Zigbee and BLE, and on the 900\,MHz band for LoRa and Z-Wave, and demonstrated significant improvement over sequential passive scanning.

Our SDR implementations should prove especially useful in overcoming the incompatibility of different protocols based on the same PHY layer.
%(in the absence of an SDR, a different network card must be used to decode packets for each protocol). 
For instance, besides Zigbee, there exist several IoT protocols based on the IEEE~802.15.4 standard, such as Thread~\cite{ThreadWhatIs} and WirelessHART~\cite{WirelessHART}. %, though they use different parameters for the PHY and MAC layers.
We expect that these protocols could readily be integrated into {\tt IoT-Scan}. 

Our SDR implementations of LoRa overcomes a significant limitation of existing network cards for this protocol. Specifically, the LoRa protocol encodes the network ID using the synchronization word (sync word) at the PHY layer. Common network cards are programmed to only receive packets containing a specific sync word and therefore cannot detect devices belonging to other LoRa networks. In contrast, our SDR implementation of a LoRa receiver gives us the opportunity to listen promiscuously to LoRa traffic. This flexibility is achieved by skipping the network address filtering enforced by the sync word, and allows us to receive all LoRa traffic regardless of the network ID. Likewise, common Z-Wave devices do not support promiscuous mode, supposedly to avoid intercepting traffic from other networks~\cite{7913565}. Our SDR implementation overrides this security-by-obscurity feature of Z-Wave.

The design of {\tt IoT-Scan} does not raise ethical issues by itself. However, like other penetration testing tools, usage of this tool does require explicit consent from the owners of the devices under test. Specifically, active scanning, while brief, may interfere with existing network traffic and delay time-sensitive communication. A major advantage of {\tt IoT-Scan} versus a tool like Nmap is that it also supports a passive scanning mode, which does not generate traffic. Nevertheless, passive scanning also requires owner consent as this may otherwise expose private information or identifiable addresses. Public releases of data logs of IoT traffic captured with {\tt IoT-Scan} must therefore be properly anonymized.  

This paper opens several avenues for future work. First, one could explore FPGA implementations of  {\tt IoT-Scan} to increase the number of channels and protocols that can be decoded in parallel and further speed up the discovery of IoT devices. While this should yield useful performance improvements, we expect that such implementations would still rely on the algorithms introduced in Section~\ref{sec:algorithms}. Another interesting research avenue lies in the design of active scanning methods for LoRa and Z-Wave, as devices in these protocols transmit sparingly.

We plan to make our implementation of {\tt IoT-Scan} broadly available to the research community to facilitate such innovations, and more generally strengthen the visibility and security of IoT devices. Anonymized time-series data extracted from PCAP files containing recorded device traffic are available at \url{https://github.com/nislab/iot-scan/}.

\section*{Acknowledgements}
This research was supported in part by the US National Science Foundation under grants CNS-1717858, CNS-1908087, CCF-2006628, EECS-2128517, and by an Ignition Award from Boston University.
% Select the IEEEtran style
\bibliographystyle{IEEEtran}
% Include bibliography file
\bibliography{IEEEabrv,main}
\end{document}